\documentclass[pre,11pt,showpacs,tightenlines]{revtex4}

\usepackage{amsmath}
\usepackage{amsfonts}
\usepackage{graphicx}
\usepackage{bbm}

\newcommand{\average}[1]{\langle #1\rangle}
\newcommand{\laverage}[1]{\big\langle #1\big\rangle}
\newcommand{\eq}[1]{(\ref{eq:#1})}
\newcommand{\eqlabel}[1]{\label{eq:#1}}
\newcommand{\sfrac}[2]{\mbox{$\frac{#1}{#2}$}}
\newcommand{\sS}{\text{\tiny S}}
\newcommand{\sT}{\text{\tiny T}}
\newcommand{\sW}{\text{\tiny $W$}}
\newcommand{\sH}{\text{\tiny $Q$}}
\DeclareMathOperator{\im}{Im}
\DeclareMathOperator{\re}{Re}

\begin{document}

\title{Extended Heat-Fluctuation Theorems for a System with
Deterministic and Stochastic Forces}

\author{R. van Zon and E.G.D. Cohen}

\affiliation{The Rockefeller University, 1230 York Avenue, New York,
New York 10021, USA}

\date{November 26, 2003}

\begin{abstract}
Heat fluctuations over a time $\tau$ in a non-equilibrium stationary
state and in a transient state are studied for a simple system with
deterministic and stochastic components: a Brownian particle dragged
through a fluid by a harmonic potential which is moved with constant
velocity.  Using a Langevin equation, we find the exact Fourier
transform of the distribution of these fluctuations for all $\tau$.
By a saddle-point method we obtain analytical results for the inverse
Fourier transform, which, for not too small $\tau$, agree very well
with numerical results from a sampling method as well as from the fast
Fourier transform algorithm.  Due to the interaction of the
deterministic part of the motion of the particle in the mechanical
potential with the stochastic part of the motion caused by the fluid,
the conventional heat fluctuation theorem is, for {\em infinite} and
for {\em finite} $\tau$, replaced by an extended fluctuation theorem
that differs noticeably and measurably from it. In particular, for
large fluctuations, the ratio of the probability for absorption of
heat (by the particle from the fluid) to the probability to supply
heat (by the particle to the fluid) is much larger here than in the
conventional fluctuation theorem.
\end{abstract}


\maketitle

\section{Introduction}

Knowledge of the behavior of heat fluctuations has an intrinsic value,
especially for small systems such as nano-systems and bio-molecules,
where the fluctuations are relatively large, but the wide-spread
interest in fluctuation theorems (FTs) stems mostly from the fact that
although there are few general results in non-equilibrium statistical
mechanics, these theorems seem to provide some.  The conventional FTs
state that the probability distribution function (PDF) $\pi$ for the
average over a time $\tau$ of a physical quantity $A$ to have a value
$a$,
satisfies\cite{Evansetal93,Evansetal94,GallavottiCohen95a,Cohen97,Kurchan98,LebowitzSpohn99}
\begin{equation}
  \frac{\pi(\average{A}_\tau=a;\tau)}{\pi(\average{A}_\tau=-a;\tau)}
\approx \exp[a\tau].  \eqlabel{generalFT}
\end{equation}
where $\average{A}_\tau$ denotes the time average of $A$ and the
dependence of the PDF on $\tau$ has been explicitly indicated.  The
asymmetry in Eq.~\eq{generalFT} is caused by some external field that
is present in these models, such that a positive value of
$\average{A}_\tau$ can correspond to a behavior ``with the field'' and
a negative value to a behavior ``against the field''.  Note that for
large $\tau$, positive values for $a$ become much more probable than
negative ones.  To be more concrete, in dynamical systems theory, the
quantity $A$ pertains to fluctuations in the ``phase space contraction
rate'' \cite{GallavottiCohen95a}, while in stochastic systems, $A$
involves an ``action functional''\cite{LebowitzSpohn99}.  In specific
cases, both have been connected with the heat or entropy
production. Such a connection is based on the expression of the phase
space contraction rate and the action functional, respectively, which
typically have the form of a thermodynamic force times a current,
divided by the temperature of the system\cite{fnone}, i.e., the same
form as the entropy production in Irreversible Thermodynamics. It is
however not necessary to make the connection with the entropy
production and in this paper, we will simply use the heat.

There are in fact two different kinds of FTs: a transient (TFT) and a
stationary state FT (SSFT).  While the conventional SSFT only holds
for sufficiently large (strictly infinite) times, the conventional TFT
holds as an identity for all times\cite{CohenGallavotti99}.  Thus, for
the SSFT, the $\approx$ sign in Eq.~\eq{generalFT} indicates the large
$\tau$ behavior, whereas for the TFT, it can be replaced by an
equality sign\cite{fntwo}.

Apart from an (early) laboratory experiment on the
SSFT\cite{otherexperiment}, the FTs were restricted
to theoretical and simulation approaches: it was difficult to make
laboratory experiments on macroscopic systems, since the large number
of particles reduces all fluctuations enormously.  Recently, Wang {\it
et al.}\cite{Wangetal02} measured a TFT in the laboratory, by studying
the motion of a {\em single} Brownian particle dragged through water
by a laser-induced moving (confining) potential.  But while in
Ref.~\cite{Wangetal02} the entropy production (or heat) fluctuations over a
time $\tau$ were intended to be studied in a transient state, in fact
the fluctuations in the work done on the system during that time were
studied.  These differ from the heat fluctuations due to the joint
presence of the confining potential and the water, as we will explain
more clearly later in the paper.  While for this system the
conventional SSFT and TFT do hold for the total {\em work} done on the
system\cite{VanZonCohen02b}, for {\em heat} fluctuations, very
different FTs hold, in which the behavior of large heat deviations is
notably and measurably different from the conventional
FTs, as shown in a recent letter\cite{VanZonCohen03a}. 
Here we present the full theory underlying the conclusions
of Ref.~\cite{VanZonCohen03a}.

The paper is organized as follows. Sec.~\ref{Theory} contains the
basic theory. We present the model (Sec.~\ref{TheoryModel}), explain
the difference between work and heat in the model
(\ref{TheoryWorkHeat}), and treat the work-related SSFT and TFT fully
(\ref{TheoryWork}). For the heat fluctuations there is no expression
for the PDF in terms of known functions, but its Fourier transform can
be calculated exactly (\ref{TheoryHeat}). Since this has no known
exact inverse, we first calculate, in Sec.~\ref{Numerical}, the PDF of
the fluctuations using two numerical methods: a sampling method as
well as the algorithm of the fast Fourier transform to perform the
inversion of the Fourier transform.  The results of both methods agree
well with each other and point to violations of the conventional SSFT
and TFT.  Next, this is substantiated by an analytic method developed
Sec.~\ref{Analytic} based on the saddle-point method, which is shown
by a comparison with the numerical methods, to work well when $\tau$
is not too small.  In Sec.~\ref{Extension} we show how this method can be
used to obtain expressions for the extended heat FTs, both in the
limit of $\tau\to\infty$, as well as for finite times.  We conclude
with a discussion of the results in Sec.~\ref{Discussion}. At the end,
two appendices give some technical details used in the text.

\section{Basic theory}
\label{Theory}         

\subsection{Model}
\label{TheoryModel}

Theoretical descriptions of the experiment of Wang {\it et
al.\/}\cite{Wangetal02} have been given even before the experiment by
Mazonka and Jarzynski\cite{MazonkaJarzynski99}, as well as recently by
the authors\cite{VanZonCohen02b,VanZonCohen03a}. All were based on an
overdamped Langevin equation for the Brownian particle, which reads:
\begin{equation}
  \dot{\mathbf{x}}_t = - \tau_r^{-1} ({\mathbf{x}}_t - {\mathbf{x}}^*_t) + \alpha^{-1} 
{\mathbf \zeta}_t.  \eqlabel{fulleqmotion}
\end{equation}
Here ${\mathbf{x}}_t$ is the (three-dimensional) position of the Brownian
particle at time $t$, $\alpha = 6\pi\eta R$ is the (Stokes) friction
of the particle in water, $\eta$ is the shear viscosity of water and
$R$ is the radius of the particle. Furthermore, the relaxation time of
the position of the particle in Eq.~\eq{generalFT}, $\tau_r$, equals
$\alpha/k$, where $k$ is the strength of the harmonic force. This
force is derived from the potential
\begin{equation}
  U({\mathbf{x}},t)=\frac{k}{2}|{\mathbf{x}}-{\mathbf{x}}^*_t|^2,
\eqlabel{2b}
\end{equation}
where ${\mathbf{x}}_t^*$ is the position of the minimum of the harmonic
potential at time $t$, given in our case by
\begin{equation}
  {\mathbf{x}}_t^*=\mathbf{v}^* t,
\eqlabel{2c}
\end{equation}
(for $t>0$) with $\mathbf{v}^*$ the (constant) velocity of the motion of the
potential through the water.  Finally, the random force ${\mathbf
\zeta}_t$ is taken to be Gaussian and delta function correlated in
time: $\average{{\mathbf \zeta}_t} =0$ and $\average{{\mathbf
\zeta}_t{\mathbf\zeta}_s}=2k_BT\alpha\delta(s-t)\mathbbm{1}$.  Here, the
brackets denote an average over realizations, $T$ is the temperature,
$k_B$ is Boltzmann's constant and $\mathbbm{1}$ denotes the $3\times3$
identity matrix.

In the rest of this paper, we will use a reduced set of units. For our
time unit we will use $\tau_r$: thus wherever $t$ appears one should
read $t/\tau_r$; for our energy unit, we choose $k_BT$ and
for our length unit, we use the width of the equilibrium
PDF of ${\mathbf x}_t$ in the potential, i.e. $\sqrt{k_B
T/k}$. This change of units has been carried through consistently in the
paper by simply setting $\alpha=1$, $k=1$ and $k_BT=1$. Then the
Langevin equation above in Eq.~\eq{fulleqmotion} takes the simple form
\begin{equation}
  \dot{\mathbf{x}}_t = - ({\mathbf{x}}_t-\mathbf{v}^* t) + {\mathbf \zeta}_t,
\eqlabel{eqmotion}
\end{equation}
where $\zeta_t$ satisfies
\begin{align}
  \average{{\mathbf \zeta}_t}&=0,
\eqlabel{avzeta}
\\
   \average{{\mathbf \zeta}_t{\mathbf \zeta}_s}&=2\delta(t-s)\mathbbm{1}.
\eqlabel{avzetazeta}
\end{align}

In Ref.~\cite{VanZonCohen02b}, the distribution of ${\mathbf{x}}_t$ and its time
autocorrelation function were determined, under the assumption that
the particle has been in the potential from time $t=-\infty$ up to
$t=0$ and that during that time the potential did not move, so that
the initial PDF at time $t=0$ is the equilibrium one. We found for
the PDF of the position ${\mathbf{x}}$ of the Brownian particle at time $t$:
\begin{equation}
  \rho({\mathbf{x}};t) = (2\pi)^{-3/2}
  e^{-|{\mathbf{x}}-\mathbf{v}^*(t-1+e^{-t})|^2/2}.
\eqlabel{xpdf}
\end{equation}
The resulting average position $\average{{\mathbf{x}}_t}$ is therefore
\begin{equation}
  \average{{\mathbf{x}}_t} = \mathbf{v}^* (t-1+e^{-t}).
\eqlabel{avX}
\end{equation}
Equations \eq{xpdf} and \eq{avX} show that the system is not in
equilibrium, but that, after a transient time period of order $1$
(after which one may neglect the term $e^{-t}$) the system becomes
stationary in a co-moving frame --- the PDF shifts to the right
with a constant velocity $\mathbf{v}^*$. The mean position of the particle
then becomes $\mathbf{v}^*(t-1)$, which is not equal to the position of the
minimum of the potential, ${\mathbf{x}}^*_t=\mathbf{v}^* t$. The difference 
$-\mathbf{v}^*$ shows we are in a stationary state where
the average harmonic force $-(\average{{\mathbf{x}}_t}-\mathbf{v}^*t)=\mathbf{v}^*$ balances
the friction force $-\mathbf{v}^*$.

Below, we will need the correlations of the position at different
times.  For the time-autocorrelation function of ${\mathbf{x}}_t$,
Ref.~\cite{VanZonCohen02b} gives in current units:
\begin{equation}
  \average{{\mathbf{x}}_{t_2}{\mathbf{x}}_{t_1}}
  -\average{{\mathbf{x}}_{t_2}}  \average{{\mathbf{x}}_{t_1}}
  = e^{-|t_2-t_1|}\mathbbm{1}.
\eqlabel{avXtX}
\end{equation}

\subsection{Work versus heat fluctuations} 
\label{TheoryWorkHeat}

We first consider the fluctuations in the total work done on the
system.  This work is the total amount of energy put into the
system. By the system, we mean here the particle in the potential $U$
plus the water. At any time, the harmonic potential exerts a force
${\mathbf F}_{t}=-({\mathbf{x}}_t-\mathbf{v}^*t)$ on the particle. Consequently, the
particle exerts a reaction force $-{\mathbf F}_t$ on the potential.
The potential has to be kept moving at a fixed velocity $\mathbf{v}^*$, for
which an external force on it of magnitude ${\mathbf F}_t$ is required
to balance the reaction force. Hence, the fluctuating work $W_\tau$
done on the system over a time interval from $t$ to $t+\tau$ is given
by
\begin{equation}
  W_\tau = \int_t^{t+\tau} \!\!d t'\, \mathbf{v}^*\cdot{\mathbf F}_{t'}
= \int_t^{t+\tau} \!\!d t'\, \mathbf{v}^*\cdot[-({\mathbf{x}}_{t'}-\mathbf{v}^*t')].
\eqlabel{Wdef}
\end{equation}

We next consider the heat $Q_\tau$ and its fluctuations, which in
contrast to the total work $W_\tau$, is only that part of the work
which goes into the fluid, while the other part of the total work goes
into the potential energy of the particle in the harmonic
potential. Therefore
\begin{equation}
  Q_\tau = W_\tau - \Delta U_\tau,
\eqlabel{Qdef}
\end{equation}
where
\begin{equation}
  \Delta U_\tau = U({\mathbf{x}}_{t+\tau},t+\tau)-U({\mathbf{x}}_t,t).
\eqlabel{Udef}
\end{equation}
The potential energy $U$ was given in Eq.~\eq{2b}\cite{fnthree}. Note
that in Eqs.~\eq{Wdef}--\eq{Udef} we suppressed the $t$ dependence.

Before we discuss the fluctuations in the work and the heat, we want
to say a few words on their average behavior. Using Eqs.~\eq{xpdf} and
\eq{Wdef}, we find that the average work in a time interval from $t$
to $t+\tau$ is given by
\begin{align}
  \average{W_\tau}
                   &=  w \left[\tau-(1-e^{-\tau})e^{-t}\right],
\eqlabel{avW}
\end{align}
where
\begin{equation}
  w \equiv |\mathbf{v}^*|^2
\eqlabel{wdef}
\end{equation}
{}From Eq.~\eq{avW}, one sees that after a transient time $t$ of $O(1)$,
the second term within the square brackets can be neglected. Then, i.e.,
in the stationary state, the average work is $w \tau$, which means
that the rate of work $w$ done on the system is constant.  To calculate
the average heat, we use Eq.~\eq{Qdef} and consider first the average
potential energy
\begin{align}
 \average{U({\mathbf{x}}_t,t)} &= \int \!\!d {\mathbf{x}}\, \rho({\mathbf{x}},t)
 \,\sfrac12|{\mathbf{x}}-\mathbf{v}^*t|^2
= \sfrac{3}{2} + \sfrac{1}{2}w (1-e^{-t})^2,
\eqlabel{avU}
\end{align}
[by Eqs.~\eq{xpdf} and \eq{wdef}], from which we find, by Eq.~\eq{Udef}, that
$\average{\Delta U_\tau}= w\left[ (1-e^{-\tau})e^{-t} -
\sfrac12(1-e^{-2\tau})e^{-2t} \right]$. Combining this with Eqs.~\eq{Qdef}
and \eq{avW}, one obtains
\begin{equation}
  \average{Q_\tau} =
  w\left[\tau+\sfrac12(1-e^{-2\tau})e^{-2t}\right].
\eqlabel{avQ}
\end{equation}
From this equation, we see that for the heat too, after a transient
time of $O(1)$, the rate of heat production becomes a
constant. Furthermore, this constant is the same as that for the work
in Eq.~\eq{avW}. So in the stationary state, {\it on average\/} all work
is converted into heat. One might therefore be tempted to identify
heat and work, at least in the stationary state. However, it will
become clear that such identification is {\em not possible} for the
{\em fluctuations} in work and heat.

\subsection{Distribution of work fluctuations}
\label{TheoryWork}

Here we will briefly summarize the results for the work fluctuations
from Ref.~\cite{VanZonCohen02b} needed in this paper.
The work
in Eq.~\eq{Wdef} is defined as an integral over the path the particle.
Since $W_\tau$ is linear in the path ${\mathbf{x}}_t$, and the ${\mathbf{x}}_t$ are
Gaussian distributed, it follows that $W_\tau$ is Gaussian distributed
as well. Hence, only the first two moments of the PDF need to be
considered.

For the work-related TFT, let us first consider the PDF for work done
on the system during a {\em transient state} of time $\tau$ , i.e.,
for an ensemble of transient trajectories of duration $\tau$, starting
in equilibrium at $t=0$\cite{Evansetal94,CohenGallavotti99}. The first
moment can then be found from Eq.~\eq{avW}, with $t=0$,
\begin{equation}
  \average{W_\tau}_\sT = w \left(\tau - 1+e^{-\tau}\right),
\eqlabel{avWT}
\end{equation}
where subscript T denotes that the average is over transient
trajectories. The second moment can be calculated as in
Ref.~\cite{VanZonCohen02b}, using the relations in Eqs.~\eq{xpdf}--\eq{Wdef}, to
be
\begin{equation}
\laverage{[W_\tau-\average{W_\tau}]^2}_\sT = 2 w \left(\tau -
  1+e^{-\tau}\right).
\eqlabel{varWT}
\end{equation}
This is exactly twice the average in Eq.~\eq{avWT}, which turns out to be
crucial. Given the mean $m$ and the variance $v$ of a general Gaussian
distributed variable $s$, its PDF is given by $P(s) =
{\exp[{-(s-m)^2/2v}]}/{\sqrt{2\pi v}}$, hence
\begin{equation}
  \frac{P(s)}{P(-s)} = e^{2ms/v}.
\eqlabel{GaussianFT}
\end{equation}
If the variance is twice the mean, $v=2m$, the right-hand side
(r.h.s.) of this equation becomes $e^s$, which is of a similar form as
the FT in Eq.~\eq{generalFT}.  To apply this equation to the work
fluctuations in a convenient way, we define $p$ as the average rate of
work in time $\tau$, $W_\tau/\tau$, scaled by the average rate $w$ of
work in the stationary state:
\begin{equation}
  p  \equiv \frac{W_\tau}{w\tau}.
\eqlabel{defwt}
\end{equation}
The transient PDF $P^\sW_\sT(W_\tau;\tau)$ of $W_\tau$ and the
transient PDF $\pi^\sW_\sT(p;\tau)$ of $p$ are related by a constant
Jacobian:
\begin{equation}
  \pi^\sW_\sT(p;\tau) = w\tau P^\sW_\sT(W_\tau;\tau).
\eqlabel{jacpi}
\end{equation}
Superscript $W$ denotes that these are PDFs related
to work.  Using Eq.~\eq{GaussianFT} and the fact that for $s=W_\tau$ the
variance is twice the mean, one finds the conventional TFT:
\begin{equation}
  \frac{\pi^\sW_\sT(p;\tau)}{\pi^\sW_\sT(-p;\tau)} = \exp[W_\tau] =
  \exp[w\tau p].
\eqlabel{WTFT}
\end{equation}
Note that the Jacobian drops out on the left hand side of this
equation.  Equation \eq{WTFT} coincides with Eq.~\eq{generalFT}, with
$\average{A}_\tau=wp$, and shows that the TFT holds for work
fluctuations. In fact, it holds exactly (``as an
identity''\cite{CohenGallavotti99}) for all times $\tau$.

For the work-related SSFT, one should in principle look at a single
(half-infinite) trajectory in the stationary state, and consider the
statistics of work done in time intervals of length $\tau$ along that
trajectory. However, because of its stochastic nature, our system
behaves ergodically and the distribution of the initial points of
these intervals is simply the stationary one.  Thus, the desired
statistics can be determined by considering a single interval $\tau$
of which the initial point is sampled from the stationary state. In
addition, as shown in the previous sections, the system reaches a
stationary state as $t\to\infty$, in an exponential fashion.  Hence we
can generate the sampling of the initial point of the interval by
considering the interval $[t,t+\tau]$ for $t\to\infty$.  In that
limit, Eq.~\eq{avW} gives
\begin{equation}
  \average{W_\tau}_\sS = w\tau.
\eqlabel{avWS}
\end{equation}
The subscript S denotes that the average is to be taken over
trajectories in the stationary state, in the $t\to\infty$ limit just
explained. The variance can be computed similarly, and the result
is\cite{VanZonCohen02b}
\begin{equation}
\laverage{[W_\tau-\average{W_\tau}]^2}_\sS = 2 w \left(\tau -
  1+e^{-\tau}\right). 
\eqlabel{varWS}
\end{equation}
Note that now the variance in Eq.~\eq{varWS} is not equal to twice the
mean in Eq.~\eq{avWS}, except when $\tau\to\infty$. We therefore expect
that only for $\tau\to\infty$, we have
\begin{equation}
  \frac{\pi^\sW_\sS(p;\tau)}{\pi^\sW_\sS(-p;\tau)} \approx \exp[W_\tau] =
  \exp[w\tau p],
\eqlabel{expectWSSFT}
\end{equation}
where $\pi_\sS^\sW$ is the PDF of the time-averaged scaled work $p$ in
the stationary state. Since the r.h.s. goes to infinity in the
$\tau\to\infty$ limit, Eq.~\eq{expectWSSFT} is not a well-defined
result. A more precise form of the SSFT is:
\begin{equation}
  \lim_{\tau\to\infty} f^\sW_\sS(p;\tau) = p,
\eqlabel{WSSFT}
\end{equation}
where one defines the \emph{work fluctuation function} as
\begin{equation}
  f^\sW_\sS(p;\tau) \equiv \frac{1}{w\tau}\ln\left[
    \frac{\pi^\sW_\sS(p;\tau)}{\pi^\sW_\sS(-p;\tau)}
    \right].
\eqlabel{defF}
\end{equation}
Indeed, using Eqs.~\eq{GaussianFT}, \eq{avWS} and \eq{varWS}, Eq.~\eq{WSSFT}
follows, i.e., the conventional SSFT for the work fluctuations
holds\cite{VanZonCohen02b}.

\subsection{Distribution of heat fluctuations}
\label{TheoryHeat}

The difficulty in considering the heat $Q_\tau$ instead of the work $W_\tau$
lies herein, that it is not linear in the position of the Brownian
particle, through the contribution of $U$ in Eqs.~\eq{Qdef}--\eq{Udef},
which according to Eq.~\eq{2b} is quadratic in $\mathbf x_t$.  As a
result, the PDF of $Q_\tau$, denoted by $P^\sH_\sT$ in the transient
state and $P^\sH_\sS$ in the stationary state (with superscript $Q$
for heat), will not be Gaussian and it does not suffice to calculate
their first two moments.  Nonetheless, the Fourier transforms of these
PDFs,
\begin{equation}
\hat P_{j}(q;\tau) \equiv \int_{-\infty}^{\infty}\!\!d Q_\tau\, e^{iqQ_\tau} 
	P_{j}^\sH(Q_\tau;\tau),
\eqlabel{defFourier}
\end{equation} 
{\em can} be explicitly calculated, as will be shown below. Here, $j$
stands either for T or for S, and this notation will be used
throughout the paper.

Consider first the transient case. We start by considering
the joint PDF $P^*_{\sT}(W_\tau,{\Delta\mathbf{x}}_1,{\Delta\mathbf{x}}_2;\tau)$ of $W_\tau$, ${\Delta\mathbf{x}}_1$ and
${\Delta\mathbf{x}}_2$. Here, ${\Delta\mathbf{x}}_1$ and ${\Delta\mathbf{x}}_2$ are the positions of the particle
relative to the position of the minimum of the potential $U$ at times
$0$ and $\tau$, respectively, i.e.,
\begin{align}
  {\Delta\mathbf{x}}_1&\equiv {\mathbf{x}}_0,
\eqlabel{defdx1T}\\
  {\Delta\mathbf{x}}_2&\equiv {\mathbf{x}}_\tau - \mathbf{v}^*\tau. 
\eqlabel{defdx2T}
\end{align}
Because $W_\tau$, ${\Delta\mathbf{x}}_1$ and ${\Delta\mathbf{x}}_2$ are all linear in the path ${\mathbf{x}}_t$,
$P^*_{\sT}$ is Gaussian, so that only its mean and variance are
needed.  Using Eqs.~\eq{2b}, \eq{Qdef} and \eq{Udef}, the transient PDF
$P^\sH_{\sT}$ of $Q_\tau$ is related to $P^*_{\sT}(W_\tau,{\Delta\mathbf{x}}_1,{\Delta\mathbf{x}}_2;\tau)$ by
\begin{align}
	P_{\sT}^\sH(Q_\tau;\tau) =& \iiint dW_\tau d{\Delta\mathbf{x}}_1 d{\Delta\mathbf{x}}_2 \,
	P^*_{\sT}(W_\tau,{\Delta\mathbf{x}}_1,{\Delta\mathbf{x}}_2;\tau) 
\nonumber\\&\times
	\delta\left(Q_\tau-W_\tau+\sfrac{1}{2}
	[|{\Delta\mathbf{x}}_2|^2-|{\Delta\mathbf{x}}_1|^2]\right).  
\eqlabel{PHT}
\end{align} 
The same can be done for the stationary case.  The PDF
$P^*_{\sS}(W_\tau,{\Delta\mathbf{x}}_1,{\Delta\mathbf{x}}_2;\tau)$ is defined as the infinite time 
limit of the joint PDF of $W_\tau$, and the relative positions ${\Delta\mathbf{x}}_1$
and ${\Delta\mathbf{x}}_2$ defined by
\begin{align}
  {\Delta\mathbf{x}}_1 &\equiv {\mathbf{x}}_t - \mathbf{v}^*t
\eqlabel{defdx1SS}\\
  {\Delta\mathbf{x}}_2 &\equiv {\mathbf{x}}_{t+\tau} - \mathbf{v}^*(t+\tau).
\eqlabel{defdx2SS}
\end{align}
Like $P^*_\sT$, $P^*_\sS$ is Gaussian. From $P^*_\sS$, the stationary
state PDF of $Q_\tau$ can be found from
\begin{align}
	P_{\sS}^\sH(Q_\tau;\tau) =& \iiint dW_\tau d{\Delta\mathbf{x}}_1 d{\Delta\mathbf{x}}_2\, 
	P^*_{\sS}(W_\tau,{\Delta\mathbf{x}}_1,{\Delta\mathbf{x}}_2;\tau) 
\nonumber\\&\times
	\delta\left(Q_\tau-W_\tau+\sfrac{1}{2}
	[|{\Delta\mathbf{x}}_2|^2-|{\Delta\mathbf{x}}_1|^2]\right).
\eqlabel{PHS} 
\end{align} 
The mean and variance of $P^*_{j}$ in Eqs.~\eq{PHT} and \eq{PHS} are
calculated in App.~\ref{appA}.

Equations \eq{PHT} and \eq{PHS} cannot be integrated explicitly, due
to the quadratic nature of $\Delta U_\tau$, but their Fourier transforms can be
obtained, as is seen when one combines Eq.~\eq{PHT} or \eq{PHS} with
Eq.~\eq{defFourier} to yield
\begin{align} 
	\hat P_{j}(q;\tau) =& \iiint dW_\tau d{\Delta\mathbf{x}}_1 d{\Delta\mathbf{x}}_2\, 
	P^*_{j}(W_\tau,{\Delta\mathbf{x}}_1,{\Delta\mathbf{x}}_2;\tau) 
	\,e^{iq[W_\tau-(|{\Delta\mathbf{x}}_2|^2-|{\Delta\mathbf{x}}_1|^2)/2]}.
\eqlabel{Fourierint}
\end{align} 
Because $P^*_{j}$ and $e^{iq[W_\tau- (|{\Delta\mathbf{x}}_2|^2-|{\Delta\mathbf{x}}_1|^2)/2]}$ are both
Gaussian, the integrals in Eq.~\eq{Fourierint} can be explicitly
performed. The details of the calculation are given in
App.~\ref{appB}, with the result
\begin{equation}
  \hat P_{j}(q;\tau) =
  \frac{
   e^{
wq(i-q)\left[\tau-\frac{[1-e^{-\tau}][\Delta_j+(\Delta_j/2+2q^2)(1-e^{-\tau})]}
                          {1+(1-e^{-2\tau})q^2}\right]   }
  }
  {[1+(1-e^{-2\tau})q^2]^{3/2}},
\eqlabel{hpt}
\end{equation}
where for $j=\mathrm T$, $\Delta_\sT=1$, while for $j=\mathrm S$,
$\Delta_\sS=0$.  The PDFs $P^\sH_j$ are related to the $\hat
P_j(q;\tau)$ by the inverse Fourier transform
\begin{equation}
  P^\sH_j(Q_\tau;\tau) = \frac1{2\pi}\int_{-\infty}^\infty \!\!dq\,
  e^{-iqQ_\tau} \hat P_j(q;\tau).
\eqlabel{inverse}
\end{equation}
Before we proceed with the evaluation of $P_{j}^\sH(Q_\tau;\tau)$ from
Eqs.~\eq{hpt} and \eq{inverse}, we note that if the conventional FT were
to hold exactly, i.e., [cf. Eq.~\eq{generalFT} with
$\average{A}_\tau=Q_\tau/\tau$]
\begin{equation}
	\frac{P^\sH_{j}(Q_\tau;\tau)} 
	{P^\sH_{j}(-Q_\tau;\tau)} = e^{Q_\tau},
\eqlabel{HFT}
\end{equation}
then, by the definition \eq{defFourier}, we would have
\begin{equation}
	\hat P_{j}(q;\tau) = \hat P_{j}(i-q;\tau).
\eqlabel{TFTq}
\end{equation}
If Eq.~\eq{TFTq} does not hold, the fluctuation theorem in Eq.~\eq{HFT} cannot
hold exactly for all $\tau$.  Neither $\hat P_{\sT}(q;\tau)$ nor $\hat
P_{\sS}(q;\tau)$ in Eq.~\eq{hpt} satisfy Eq.~\eq{TFTq}. That means we can rule
out the possibility of a TFT or a SSFT for heat fluctuations that
holds as an identity for all $\tau$.

The first term in the exponent in Eq.~\eq{hpt} does have the symmetry in
Eq.~\eq{TFTq}, and this term is the only one that grows with the time
$\tau$. However, one cannot conclude from this that the FTs hold for
large $\tau$, due to the singularities in Eq.~\eq{hpt}.  In the rest of
the paper, we will be concerned with whether Eq.~\eq{HFT} holds, both for
transient and stationary states, for $\tau\to\infty$, or, if it does
not, what the deviations from this behavior are. To be precise,
similar to the treatment in Sec.~\ref{TheoryWork}, we define the scaled
heat fluctuations --- denoted by the same symbol $p$ as scaled work
fluctuations --- as 
\begin{equation}
  p  \equiv \frac{Q_\tau}{w\tau},
\eqlabel{pdef}
\end{equation}
[cf. Eq.~\eq{defwt} and the remark below Eq.~\eq{avQ}]. Its PDF is 
\begin{align}
  \pi_{j}^\sH(p;\tau)&=w\tau P^\sH_j(w\tau p\,;\,\tau)
\eqlabel{piH}
\\
&=\frac{w\tau}{2\pi}\int_{-\infty}^\infty\!\!dq\, e^{-iqw\tau p}\hat P_j(q;\tau)
\eqlabel{piH2}
\end{align}
[cf.\ Eq.~\eq{jacpi}], where Eq.~\eq{inverse} was used. Furthermore, we define the \emph{heat
fluctuation function} [cf.\ Eq.~\eq{expectWSSFT}]
\begin{equation}
  f^\sH_{j}(p;\tau) \equiv \frac{1}{w\tau}\ln\left[
    \frac{\pi^\sH_{j}(p;\tau)}{\pi^\sH_{j}(-p;\tau)}
    \right].
\eqlabel{defFH}
\end{equation}
and we investigate whether [cf.\ Eq.~\eq{defF}]
\begin{equation}
  \lim_{\tau\to\infty} f^\sH_{j}(p;\tau) = p.
\eqlabel{preciseHFT}
\end{equation}
as is required for the conventional FT.

\section{Numerical Approach}
\label{Numerical}           

\begin{figure}[t]
\centerline{\includegraphics[width=0.75\textwidth]{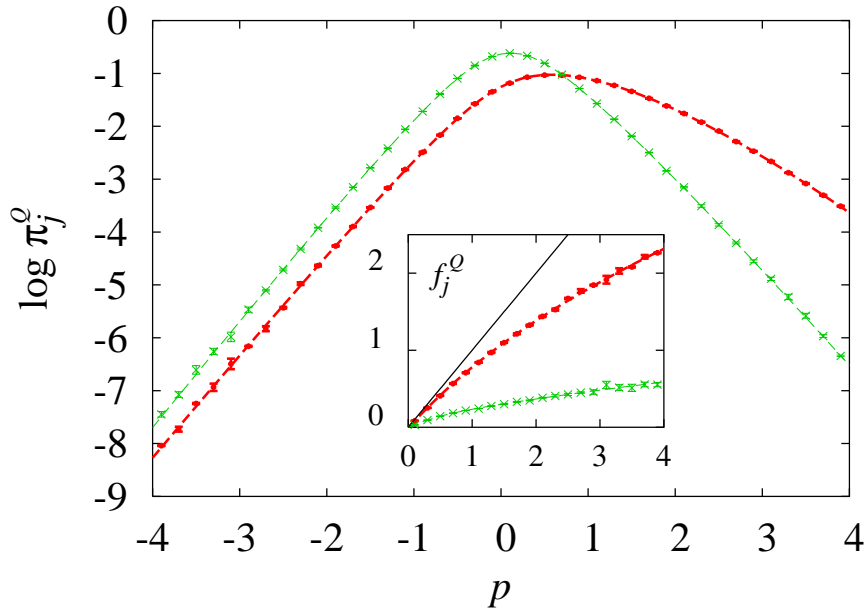}}
\caption{The PDFs $\pi^\sH_j$ obtained by two numerical methods, for
  $w=2.0$ and $\tau=1.0$. Sampling method results ($N_s=2,000,000$,
  $\delta p=0.2$) are shown as crosses for $\pi^\sH_\sT$ and
  as  dots for $\pi^\sH_\sS$, respectively. Fast Fourier
  Transform results ($\delta q=4\cdot10^{-4}$, $q_{max}=200$) are shown
  as a thin dashed  line for $\pi^\sH_\sT$ and as a bold dashed
  line for $\pi^\sH_\sS$. The inset shows the corresponding
  fluctuation functions $f^\sH_j$ [Eq.~\eq{defFH}] and a
  solid straight line indicating the prediction of the
  conventional FT.}\label{fig:1}
\end{figure}

Since no exact Fourier inverse of Eq.~\eq{hpt} is known, we first
treat the problem of finding the PDF of heat fluctuations numerically.

We used two numerical methods. The first is a sampling method which
starts from the expressions in Eqs.~\eq{PHT} and \eq{PHS} that give
the $P_j^\sH$ in terms of the $P^*_j$. From the Gaussian PDF $P^*_{j}$
with the mean and variance calculated in App.~\ref{appA}, one draws
many sets of values
$(W_\tau,{\Delta\mathbf{x}}_1,{\Delta\mathbf{x}}_2)$, using a random
number generator\cite{NumericalRecipes}. From each set, using
Eqs.~\eq{Qdef}, \eq{Udef} and \eq{pdef}, $p = (W_\tau-\Delta
U_\tau)/w\tau$ is calculated. From these $p$ values we build a
histogram by constructing bins of width $\delta p$ in an interval
$[-p_{max},p_{max}]$ and counting how many of the values fall into
each bin. If the number of sample points $N_{s}$ is large and $\delta
p$ is small, the histogram gives a good approximation for the PDF
$\pi^\sH_j$. As a simple error estimate, one performs this procedure a
few times with varying random seeds, and determines the spread in the
results.

\begin{figure}[t]
\centerline{\includegraphics[width=0.75\textwidth]{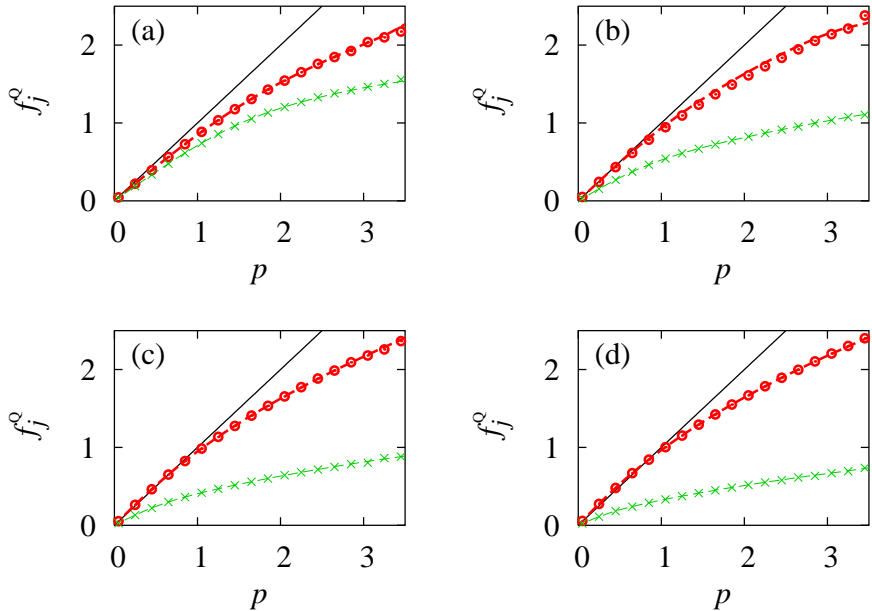}}
\caption{Numerical results for the fluctuation functions
  $f_j^\sH$ for the stationary ($j=\rm S$) and the transient case
  ($j=\rm T$), for (a) $w=1, \tau=4$, (b) $w=2, \tau=2.5$, (c) $w=3,
  \tau=1.3333$ and (d) $w=4, \tau=1$.  Sampling method results
  ($N_{s}=1.2\cdot 10^{9}$, $\delta p=0.2$) are shown as thin
  crosses for $f_\sT^\sH$ and as bold dots for $f_\sS^\sH$. Fast
  Fourier transform results ($\delta q=4\cdot10^{-4}$, $q_{max}=200$)
  are shown as the dashed thin  curves for $f_\sT^\sH$ and as
  dashed bold curves for $f_\sS^\sH$. The solid straight
  line indicates the prediction of the conventional FT.}\label{fig:2}
\end{figure}

The second numerical method is the standard {\em Fast Fourier
Transform} algorithm\cite{NumericalRecipes} applied to the inverse
Fourier transform. This algorithm takes a discrete set of values of
the function $\hat P_j(q)$ for $q$ in an interval $[-q_{max},q_{max}]$
with equal spacing $\delta q$ between the values, and returns values
of the (inverse) transform, i.e., of the function $P^\sH_j$, on a
reciprocal grid of values for $Q_\tau$.  If $\delta q$ is small
enough, and $q_{max}$ is large enough, this can be used to obtain a
good approximation for the (inverse) Fourier transform. An error
estimate can be found by varying $\delta q$ and $q_{max}$ and
observing to what extent the results have converged. In all plots in
this paper, these errors would be unobservable and are therefore not
plotted.  Note that once $P^\sH_j$ is known, by Eq.~\eq{piH} we can
obtain $\pi^\sH_j$ as well.

The results of these two methods for the PDFs $\pi^\sH_\sT$ and
$\pi^\sH_\sS$ for the parameter values $w=2.0$ and $\tau=1.0$ have
been plotted in Fig.~\ref{fig:1} as a function of $p$, and the
corresponding fluctuation functions $f^\sH_\sT$ and $f^\sH_\sS$ are
plotted in the inset in that figure.  The two numerical methods agree
very well, but do not agree at all with the straight line with slope
$1$ of the conventional FT [i.e., $p$ cf. Eq.~\eq{preciseHFT}], which
is also drawn in the inset in Fig.~\ref{fig:1}, neither in the
stationary nor in the transient case.

To investigate the discrepancies with the conventional FTs further, we
have explored a range of values for the parameters $w$ and $\tau$,
using the two numerical methods. Some results are shown in
Fig.~\ref{fig:2}.  One sees that for large values for $p$, deviations
of $f_j^{\sH}(p;\tau)$ from the straight line $p$ of the conventional
FT are generic.

\begin{figure}[t]
\centerline{\includegraphics[width=0.75\textwidth]{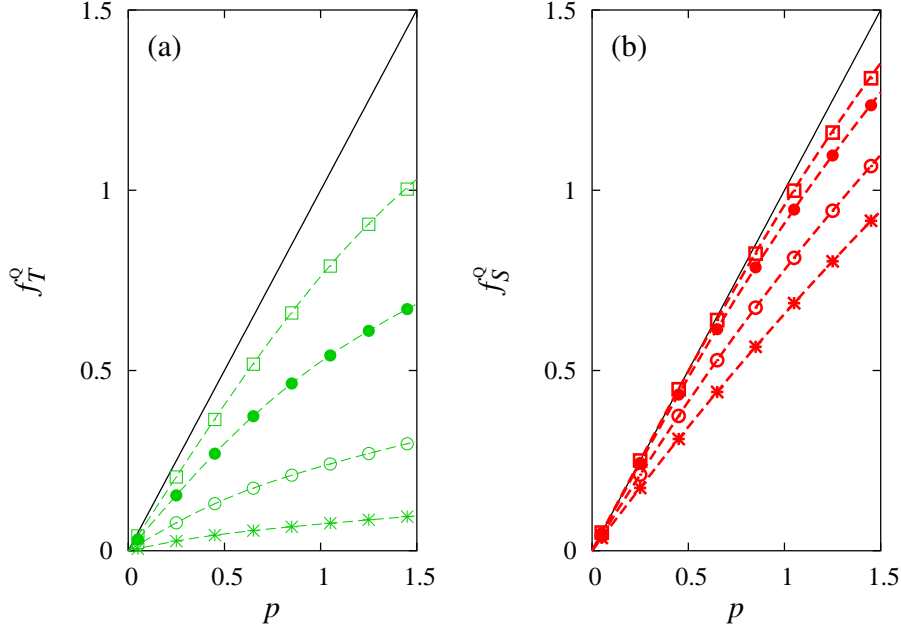}}
\caption{Numerical results for the $\tau$ dependence of the
  fluctuation functions $f_j^\sH$ for $w=2.0$ for (a) the
  transient ($j=\rm T$) and (b) the stationary case ($j=\rm S$).
  Sampling results ($N_{s}=1.2\cdot 10^{9}$, $\delta p=0.2$) are shown
  as symbols: ($*$) $\tau=0.5$, ($\circ$) $\tau=1$, ($\bullet$)
  $\tau=2$, ({\scriptsize $\Box$}) $\tau=4$.  Fast Fourier Transform
  results ($\delta q=4\cdot10^{-4}$, $q_{max}=200$) are shown as the
  dashed curves (the $\tau$ values of these curves are the same as
  those of the coinciding sampling points). The solid straight
  line is the conventional FT result.}\label{fig:2bis}
\end{figure}

Furthermore, Fig.~\ref{fig:2bis} shows (for $w=2.0$, but the same
holds for other $w$ values) that for large $\tau$, the straight line
is approached for small $p$, both in the transient and the stationary
case, although the approach is considerably slower for the former.

\section{Analytic Approach --- Saddle Point Method}
\label{Analytic}           

The numerical methods are not suitable to give reliable results for
the tails of the PDFs for large $\tau$. To get around this, we
developed an asymptotic analytic approach.

The starting point of the approach is to notice that the exponent in
Eq.~\eq{hpt} grows linearly with $\tau$. Thus, the $\pi_j^\sH$, given
by Eq.~\eq{piH2}, can be written as
\begin{equation}
  \pi_j^\sH(p;\tau) = \frac{w\tau}{2\pi}\int_{-\infty}^{\infty} \!\!dq\,
                    e^{-w\tau [e_j(q;\tau)+iqp]},
\eqlabel{largeparameter}
\end{equation}
where 
\begin{equation}
  e_j(q;\tau)\equiv-\frac{1}{w\tau}\ln[\hat P_j(q;\tau)]
\eqlabel{gjdef}
\end{equation}
is of order $\tau^0$ [see Eq.~\eq{hpt}].  The presence of a large
parameter $\tau$ in Eq.~\eq{largeparameter} makes it suitable for the
saddle-point method\cite{Jeffreys}, which we will briefly describe
now.

To get a good approximation of the integral in
Eq.~\eq{largeparameter}, one considers
\begin{equation}
  h(q) \equiv  -w[e_j(q;\tau)+iqp]
\eqlabel{deff}
\end{equation}
as a function of a complex-valued $q$ (where the dependences of $h$ on
$j$, $p$ and $\tau$ are suppressed). Then, one determines its {\em
saddle point}, i.e., the complex number $q=q^*$ for which $\partial h
/\partial q=0$. Through this point $q^*$ in the complex plane lies a
\emph{path of steepest descent} $S$ along which $\im h$ is constant
and $\re h$ attains a maximum at $q^*$ \cite{Jeffreys}.  Next, one
continuously deforms the original path of integration $R$ (here the
real axis) to this path of steepest descent $S$ without crossing
singularities.  The integral over $S$ is for large $\tau$ dominated by
the (small) segment on $S$ around the saddle point $q^*$ and can be
expanded in inverse powers of $\tau$ by Taylor expanding the function
$h$ on $S$ around the saddle point. For the leading term one uses a
second order Taylor expansion and finds\cite{Jeffreys}
\begin{equation}
  \int_{-\infty}^{\infty} \!\!dq\, e^{\tau h(q)} =
  \sqrt{\frac{2\pi}{\tau|h''(q^*)|}}\,e^{\tau h(q^*)+i\alpha} [1+O(\tau^{-1})].
\eqlabel{saddlepointmethod}
\end{equation}
Here, $\alpha$ is the angle between the direction of the path $S$ when
it traverses the saddle point, and the real axis.  Furthermore,
$h''(q^*)$ is the second derivative of $h$ with respect to $q$ at the
saddle point $q^*$, which is assumed to be non-zero. The correction
term in Eq.~\eq{saddlepointmethod} is $O(\tau^{-1})$, though that
strictly applies only when the function $h$ does not depend on
$\tau$. In our case it does, and we will see that consequently the
correction terms can then be $O(\tau^{-1/2})$, although the leading
behavior in Eq.~\eq{saddlepointmethod} is not affected.

The form of $h(q)$ in Eq.~\eq{deff} has the following simplifying
consequences for the expression in Eq.~\eq{saddlepointmethod}.
Firstly, the equation for the saddle point, $\left.\partial h/\partial
q\right|_{q=q^*}=0$, becomes
\begin{equation}
  e'_j(q^*;\tau) = -i p,
\eqlabel{saddlepoint}
\end{equation}
where $e'_j$ is the derivative of $e_j$ with respect to $q$.
Secondly, $h''(q^*)$ in Eq.~\eq{saddlepointmethod} is equal to
$-we''_j(q^*)$. This second derivative can be further simplified by
noticing that Eq.~\eq{saddlepoint} holds for all $p$, so we can take
the derivative with respect to $p$ on both sides to find
$e''_j(q^*;\tau)({\partial q^*}/{\partial p}) = -i$.  Hence
\begin{equation}
  |h''(q^*)| = w|e''_j(q^*;\tau)|=w\left|\frac{\partial q^*}{\partial
   p}\right|^{-1}.
\eqlabel{fpp}
\end{equation}
Thirdly, the approximation is for a PDF, which should be real and
positive. Thus, if the saddle-point method is to be a consistent
calculation of a PDF, we must have
\begin{equation}
  \alpha=0.
\eqlabel{alpha}
\end{equation}
We will prove later that this is indeed the case.  Using
Eqs.~\eq{deff}, \eq{fpp} and \eq{alpha}, we apply the saddle point
approximation Eq.~\eq{saddlepointmethod} to the integral in
Eq.~\eq{largeparameter} to find
\begin{equation}
  \pi^\sH_j(p;\tau) \sim \sqrt{\frac{w\tau}{2\pi}\left|\frac{\partial
  q^*}{\partial p}\right|}\,e^{-w\tau [e_j(q^*;\tau)+ipq^*]}.
\eqlabel{spapprox}
\end{equation}
Here, the symbol $\sim$ is for convenience used to denote the
asymptotic behavior, instead of explicitly denoting the correction
terms as $O(\tau^{-1})$ as we did in Eq.~\eq{saddlepointmethod}.

As mentioned above Eq.~\eq{saddlepointmethod}, this method gives an
asymptotic expansion in inverse powers of $\tau$.  Now if $\tau$ is
large enough so that terms of relative order $\tau^{-1}$ can be
neglected, then terms which are exponentially small in $\tau$ can also
be neglected\cite{fnfive}.  Therefore, we use in the definition of the
exponents $e_j$ in Eq.~\eq{gjdef} the expression of the Fourier
transform $\hat{P}_j(q;\tau)$ in Eq.~\eq{hpt} without exponential
terms, so that\cite{fnsix}
\begin{align}
  e_j(q;\tau) &= -q(i-q)-\frac{4q^3+3\Delta_j q}{2\tau(i+q)}
                +\frac{3\ln(1+q^2)}{2w\tau}
\eqlabel{gS}
\end{align}
When substituted in Eq.~\eq{saddlepoint}, we obtain a fourth order
polynomial equation for the saddle points, so that there are four
saddle points.  To know which of the four saddle points to use, one
needs to determine their location, the paths of steepest descent that
traverse them, and to see if our initial integration line (the real
axis) can be deformed to lie on one of these paths without crossing
any singularities.

\begin{figure}[t]
\centerline{\includegraphics[width=0.7\textwidth]{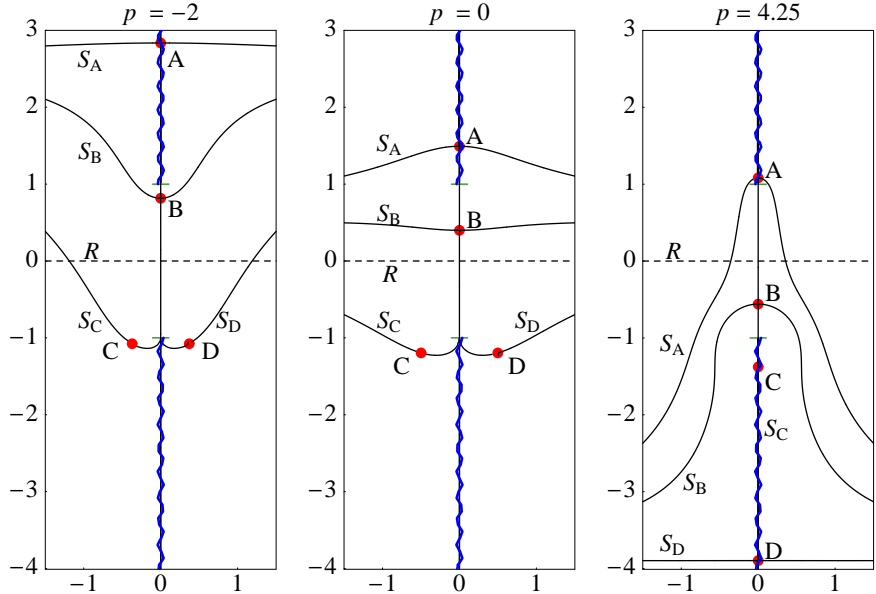}}
\caption{Geometry in the complex $q$-plane of the saddle points,
  singularities and paths of steepest descent of $h(q)$ [Eqs.~\eq{deff}
  and \eq{gS}] for the stationary state [$\Delta_j=0$], for $w=1$,
  $\tau=4$, and $p=-2$, $0$, and $4.25$, respectively. In all graphs,
  solid  dots depict the position of the saddle points A, B, C
  and D. Wiggly curves give the branch cuts ending in the
  branch points at $q=\pm i$ which are indicated with a thin horizontal
  line.  The branch point at $-i$ is also a pole.  The
  dashed line $R$ gives the original line of integration, i.e., the
  real axis. Solid lines labeled $S_{\mathrm A}$, $S_{\mathrm B}$,
  $S_{\mathrm C}$ and $S_{\mathrm D}$ are paths of steepest descent
  though the saddle point, on which $\re h$ attains a
  maximum.
} \label{fig:3}
\end{figure}

For three typical values for $p$ we have depicted the four saddle
points of $h$ for the stationary state in Fig.~\ref{fig:3}, calculated
analytically by {\em Mathematica}\cite{Mathematica} which uses
Ferrari's method for solving the quartic equation\cite{fnfour}.  We
see in Fig.~\ref{fig:3} that for all $p$ there is one saddle point (A)
on the imaginary axis above $i$, one in between $-i$ and $i$ (B), and
two (C and D) who are either both purely imaginary but below $-i$, or
are complex with the same imaginary part (also below $-i$), and with
opposite real parts.  In Fig.~\ref{fig:3} there are a pole at $q=-i$,
and branch cuts which we have taken along the imaginary axis, one from
$i$ upward and one from $-i$ downward.  These cuts come from the
logarithmic term in Eq.~\eq{gS}.  Finally and most importantly,
Fig.~\ref{fig:3} shows the paths of steepest descent $S_{\mathrm A}$,
$S_{\mathrm B}$, $S_{\mathrm C}$ and $S_{\mathrm D}$ through the
saddle points A, B, C, and D. These paths were found by having {\em
Mathematica} solve for lines of constant $\im h$, where the values of
the constant for $\im h$ are chosen to match the values of $\im h$ at
the four saddle points.

Knowing now the geometry of $h$ in the complex plane, we can determine
which saddle point to use. One sees that for all $p$ the path
$S_{\mathrm B}$ through saddle point B is a possible deformation of
the real axis: each point on the real axis is simply shifted up or
down until it lies on this path and no poles or other singularities
are crossed in doing so. Strictly speaking, the deformation changes
the end points because they get shifted by an imaginary amount $i a$
[where $a$ can be shown to be $(1-p)/2$], but one can add portions
$ia$ to the integration path from $-\infty$ to $-\infty+ia$ and from
$\infty+ia$ to $\infty$, and as the integrand vanishes on these
portions, the integral taken over $R$ and over $S_{\mathrm B}$ have
the same value.  This is not true for $S_{\mathrm A}$, $S_{\mathrm C}$
and $S_{\mathrm D}$.

The graphs in Fig.~\ref{fig:3} suggest that the path $S_{\mathrm B}$
goes through B horizontally, i.e., that $\alpha=0$ as Eq.~\eq{alpha}
conjectured. To prove this, notice that for purely imaginary values
for $q$, i.e., $q=i\lambda$, $h(i\lambda)=-w[e(i\lambda;\tau)-\lambda
p]$ is real if $-1<\lambda<1$ [cf. Eq.~\eq{gS}]. Hence, the imaginary
axis from $-i$ to $i$ is a line of constant imaginary part of $h$. In
the saddle point, lines of constant imaginary part cross
perpendicularly\cite{Jeffreys}.  Thus in saddle point B, the curve of
steepest descent is seen to be perpendicular to the imaginary axis,
i.e., in the direction of the real axis, so that indeed $\alpha=0$.

We remark that the above analysis for the stationary case can be
carried through also for the transient case, with only a slight change in
the saddle points.

Our saddle-point method now consists of the following.  We use {\em
Mathematica} to determine analytically the purely imaginary solution
$q^*$ of Eqs.~\eq{saddlepoint} using \eq{gS} with $|q^*|<1$, as a
function of $p$. Given that solution, the PDF $\pi^\sH_j$ is
explicitly calculated using Eqs.~\eq{spapprox} and \eq{gS}. Once
$\pi^\sH_j$ is known, we determine the fluctuation functions $f_j^\sH$
from Eq.~\eq{defFH}. We can then investigate the FTs, even for finite
$\tau$, something which is usually not possible, but is doable in this
case because the Fourier transform is known exactly for all time.

\begin{figure}[t]
\centerline{\includegraphics[width=0.75\textwidth]{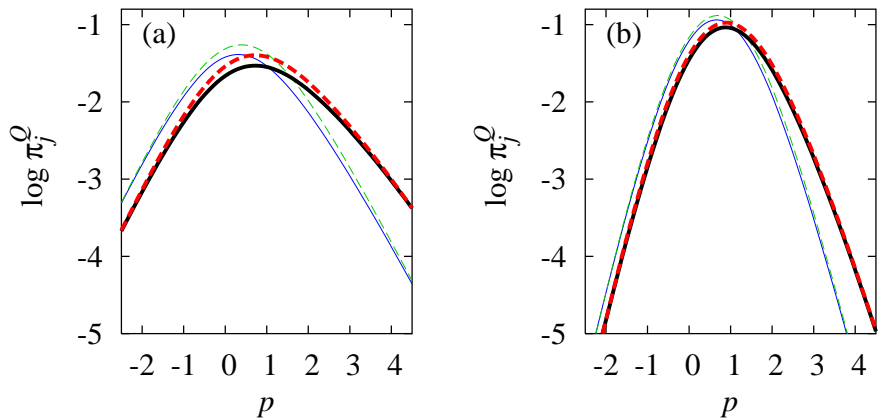}}
\caption{Comparison of the saddle-point approximation with the
  numerically obtained solution, for $w=0.5$ and (a) $\tau=2.5$ and
  (b) $\tau=5$. The solid thin line is the transient PDF
  $\pi^\sH_\sT(p;\tau)$ as calculated using the saddle-point
  approximation of Sec.~\ref{Analytic} and the dashed thin curve is
  the same PDF calculated using the numerical inverse Fourier
  transform of Sec.~\ref{Numerical} ($\delta q=2\cdot 10^{-4}$,
  $q_{max}=2^{18}\delta q = 52.4288$).  The solid bold line is the
  stationary PDF $\pi^\sH_\sS(p;\tau)$ calculated using the
  saddle-point approximation, and the bold dashed curve is that same
  PDF obtained using the numerical inverse Fourier
  transform.}\label{fig:4a}
\end{figure}

To see how good the saddle-point expression for $\pi^\sH_j$ in
Eq.~\eq{spapprox} is, we compare it with the numerical results from
Sec.~\ref{Numerical}, specifically, the numerically determined Fourier
inverse. This is shown in Fig.~\ref{fig:4a}, both for the transient
and for the stationary case. In Fig.~\ref{fig:4a}(a), we chose
$\tau=2.5$ and $w=0.5$, i.e., not too large, and we are therefore
still able to see the discrepancies between the saddle-point result
and the inverse Fourier transform for $\pi_j^\sH$, which are about
15\% near the peak, and just a few percents in the tails. In
Fig.~\ref{fig:4a}(b), we chose $\tau=5.0$, which is still not too
large, but we now see that the saddle-point method has become quite
accurate, up to about 7\% in the peak of the distribution, while in
the tails one can hardly distinguish the results from the two methods.
As $\tau$ gets larger, the results from these two methods approach
each other even more and plotting them together as in
Fig.~\ref{fig:4a} would show basically two curves on top of the other.

\begin{figure}[t]
\centerline{\includegraphics[width=0.75\textwidth]{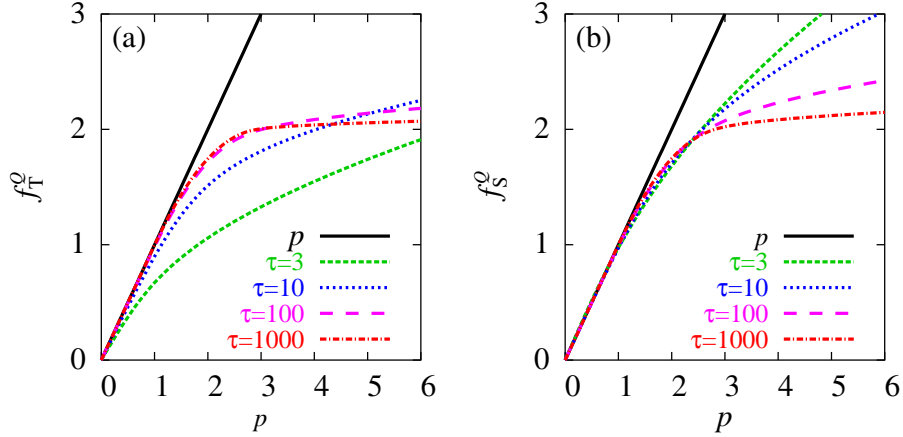}}
\caption{Results for the large $\tau$ behavior of (a)
$f_T^\sH(p;\tau)$ and (b) $f_T^\sH(p;\tau)$ for $w=1$, as found from
the saddle point approximation.  The figures show four lines
corresponding to result for $\tau=3$, $\tau=10$, $\tau=100$ and
$\tau=1000$, as well as a solid black straight line ($p$) that gives
the behavior expected from the conventional FT. }\label{fig:5}
\end{figure}

While the numerical methods in Sec.~\ref{Numerical} could give us the
large $\tau$ behavior of $f_j^\sH(p;\tau)$ only for small $p$ values,
using the saddle-point approximation, we can now determine the
behavior of $f_j^\sH(p;\tau)$ with increasing $\tau$ for the full
range of $p$ values.  The results are plotted in Fig.~\ref{fig:5} and
show that both for the transient and for the stationary case, the
fluctuation function $f_j^\sH$ approaches the conventional FT only for
small $p$ values. In contrast, for large $p$ values ($p>3$), a
completely different limit emerges as $\tau\to\infty$, one where
$f_j^\sH(p,\tau)$ appears to approach $2$, both for the transient and
for the stationary case.  The exact form of this $\tau\to\infty$ limit
of $f_j^\sH(p,\tau)$ is given by Eq.~\eq{infiniteasympFH} in
Sec.~\ref{Extension}.

\section{Analytic Expressions for Extended Fluctuation Theorems}
\label{Extension}                               

Because we do not find it satisfactory to have the analytical result
of the saddle-point method only in the memory of {\em Mathematica}, we
will determine in this section explicitly the leading asymptotic
behavior of $\pi^\sH_j$ and corrections as $\tau$ gets large, using
Eqs.~\eq{saddlepoint}, \eq{spapprox} and \eq{gS}.

To calculate the saddle-point approximation for $\pi^\sH_j$ given in
Eq.~\eq{spapprox}, we need to find the saddle-point $q^*$ first.  With
$e_j$ from Eq.~\eq{gS}, Eq.~\eq{saddlepoint} for $q^*$ takes the form
\begin{equation}
1+2iq^* +
\frac{4q^{*2}(3-2iq^*)+3\Delta_j}{2\tau(i+q^*)^2} 
-\frac{3iq^*}{w\tau(1+q^{*2})}
=p.
\eqlabel{spexplicit}
\end{equation}
This equation always has four (complex) solutions as explained below
Eq.~\eq{gS}. Among these four solutions, one is always purely
imaginary and lies between $-i$ and $i$, which is the one we need, as
it corresponds to saddle-point B (see previous section).  In order to
find this solution, we try the expansion
\begin{equation}
  q^* = q^*_0 + \frac{q^*_1}{\tau} + O(\tau^{-2}),
\eqlabel{lambdaexpanded}
\end{equation}
where $q^*_0$ is given by the solution of Eq.~\eq{spexplicit} with
$\tau\to\infty$, i.e., $1+2iq^*=p$. Hence
\begin{equation}
q^*_0=i(1-p)/2.
\eqlabel{modpzero}
\end{equation}
Since $q^*$ lies between $-i$ and $i$, the above zeroth order solution
is only appropriate if it lies in that interval, i.e., if $-1<p<3$. We
will refer to this as case (a), which will be dealt with first. The
cases (b) $p<-1$ and (c) $p>3$ will be treated later.

(a) For $-1<p<3$, Eq.~\eq{modpzero}, when substituted into
Eq.~\eq{spapprox} using $e_j$ from Eq.~\eq{gS}, yields a zeroth order
solution for $\pi^\sH_j(p;\tau)$. Because $q^*_0$ is the maximum of
$h$ in Eq.~\eq{saddlepointmethod}, there is no need to calculate
$q^*_1$, as far as the exponent $h$ is concerned\cite{fnnine}.
Furthermore, in the prefactor in Eq.~\eq{spapprox}, $\partial
q^*_0/\partial p$ is of order $1$, so adding $\tau^{-1}\partial
q^*_1/\partial p$ gives a correction of relative order $O(\tau^{-1})$,
i.e. of the same order as the correction terms in
Eq.~\eq{saddlepointmethod}.  So there is no point to work out $q^*$ to
higher orders than $q^*_0$.  Substituting the expression for $q^*_0$,
given in Eq.~\eq{modpzero}, for $q^*$ in Eq.~\eq{spapprox} and using
Eq.~\eq{gS}, yields
\begin{multline}
\pi^\sH_j(p;\tau) \sim \sqrt{\frac{16w\tau}{\pi}}\,\frac{e^{-w\left[
      \tau(1-p)^2 +\frac{2(1-p)}{3-p}\{(1-p)^2-3\Delta_j\}\right]/4}}
{[(3-p)(1+p)]^{3/2}}
\qquad\qquad\qquad
\mbox{for  $\,-1<p<3$,}
\eqlabel{piHmid}
\end{multline}
with correction terms that are relatively $O(\tau^{-1})$. 

Outside of the range $-1<p<3$, the zeroth order Ansatz in
Eq.~\eq{modpzero} fails to produce a solution for $q^*$ between $-i$
and $+i$. The explanation is that the expression for $q_0^*$ in
Eq.~\eq{modpzero} was based on neglecting the last two terms on the
left-hand side (l.h.s) of Eq.~\eq{spexplicit}, which clearly does not
work for $p>3$ or $p<-1$. However, the only way that these terms can
become comparable to the first term in that equation, is if their
denominators become $O(1)$ instead of $O(\tau)$ as they appear to
be. For this another Ansatz than Eq.~\eq{lambdaexpanded} is
necessary. As there are two different denominators in
Eq.~\eq{spexplicit}, there are two cases. In the first case, the
denominator of the last term on the l.h.s.\ of Eq.~\eq{spexplicit} is
taken to be $O(1)$, i.e., $\tau(1+q^{*2})=O(1)$, or $q^* =\pm i +
O(\tau^{-1})$. The sign ambiguity is lifted by noting that for
$q^*=-i+O(\tau^{-1})$, the second term in Eq.~\eq{spexplicit} would
diverge $\propto\tau$, so only
\begin{equation}
  q^*=+i+O(\tau^{-1})
\eqlabel{lstarp1}
\end{equation}
is correct.
In the second case, the denominator of the second term on the
l.h.s.\ of Eq.~\eq{spexplicit} is taken to be $O(1)$,
i.e. $\tau(i+q^*)^2=O(1)$, or 
\begin{equation}
q^* = -i + O(\tau^{-1/2}).
\eqlabel{lstarm1} 
\end{equation}

(b) We first consider the case $q^*=i+O(\tau^{-1})$ of Eq.~\eq{lstarp1},
which we write for convenience as\cite{fnten}
\begin{equation}
  q^*=i+\frac{ia}{w\tau}+O(\tau^{-2}).
\eqlabel{le2}
\end{equation}
Substituting Eq.~\eq{le2} into Eq.~\eq{spexplicit}, and keeping only $O(1)$
terms, one gets $-1 + \frac{3}{2a} = p$, so $a=\frac{3}{2(p+1)}$,
and
\begin{equation}
  q^*=i+\frac{3i}{2w\tau(p+1)} + O(\tau^{-2}).
\eqlabel{smallplambda}
\end{equation}
Therefore, for $p<-1$, this is an approximation of the solution
$q^*$ below $i$. Using Eqs.~\eq{spapprox}, \eq{gS} and
\eq{smallplambda}, one finds
\begin{multline}
\pi^\sH_j(p;\tau) \sim
\sqrt{\frac{w^3\tau^3|p+1|}{36\pi}}
e^{-w[-\tau p +1 -\frac34\Delta_j]+\frac{3}{2}}\,
\qquad\qquad\qquad\qquad\qquad\qquad
\mbox{ for $p<-1$.}
\eqlabel{piHlow}
\end{multline}
Again, correction terms to this are of relative $O(\tau^{-1})$.

(c) Finally, we consider the case $q^*=-i+O(\tau^{-1/2})$ of
Eq.~\eq{lstarm1} and write
$q^*=-i+\frac{ib}{\sqrt\tau}+\frac{ic}{\tau}$.  Note that in
principle, we needed to include two correction terms here to get a
similar level of approximation as in Eqs.~\eq{lambdaexpanded} and
\eq{le2}.  However, as the sum of the first two terms (i.e.,
$-i+ib/\sqrt{\tau}$) is to leading order the maximum of the exponent,
we need not calculate $c$\cite{fnnine}. Substituting
$q^*=-i+ib\tau^{-1/2}$ into Eq.~\eq{spexplicit}, gathering the $O(1)$
terms and using that $\Delta_j^2 = \Delta_j$, we find
$b=\pm(2-\Delta_j)/\sqrt{2(p-3)}$.  We need the positive solution, so
that $q^*$ is above $-i$, i.e.,
\begin{align}
  q^*&=-i+i\frac{2-\Delta_j}{\sqrt{2(p-3)\tau}} + O(\tau^{-1}),
  \eqlabel{d}
\end{align}
Note that we need $p>3$ for this to be purely imaginary, as it should
be. Combining Eqs.~\eq{spapprox}, \eq{gS} and \eq{d} gives
\begin{multline}
\pi^\sH_j(p;\tau) \sim
\sqrt{\frac{w\tau^2}{16\pi(2-\Delta_j)^2}}
\, e^{-w\left[(p-2)\tau-(2-\Delta_j)\sqrt{2(p-3)\tau}
    +\frac{40-12p+3\Delta_j(p-4)}{2(p-3)}\right]}
\qquad
\mbox{ for $p>3$,}
\eqlabel{piHhi}
\end{multline}
where $\Delta_j^2 = \Delta_j$ has been used to simplify the
expression.  Due to the different form of the expansion of $q^*$ here
compared to the cases (a) and (b), this result is valid up to
corrections of relative order $O(\tau^{-1/2})$.

Note that we now have the asymptotic forms for $\pi^\sH_j$ in the
three separate regions, $p<-1$, $-1<p<3$ and $p>3$ in the
Eqs.~\eq{piHlow}, \eq{piHmid} and \eq{piHhi} respectively.  This shows
that the behavior in the center of the PDF is Gaussian-like, while in
the tails it is exponential.

To investigate the FTs, we will now consider the fluctuation functions
$f^\sH_j$.  For that, we have to take the above asymptotic expression
for large but finite $\tau$ to calculate $f_j^\sH$ of
Eq.~\eq{defFH} (after which one can take the limit $\tau\to\infty$). In
this way, one finds for the large but finite $\tau$ behavior of
$f_j^\sH$:\cite{fneleven}
\begin{equation}
  f_j^\sH(p;\tau) = \left\{\begin{array}{l}
  \displaystyle p 
  + \frac{(8-6\Delta_j)p}{\tau(9-p^2)}
  -\frac{3\ln 
\frac{3-p^2+2p}{3-p^2-2p}
}{2w\tau}
  + O(\tau^{-2}) \\
    \hfill \mbox{ for $0\leq p<1$}\\
    \displaystyle
    p-{(1-p)^2}/{4}-\frac{\ln\tau}{w\tau} + \frac{r_j(p)}{w\tau}
    +O(\tau^{-2})\\
  \hfill \mbox{ for $1<p<3$}\\
  \displaystyle
  2+(2-\Delta_j)\sqrt{\frac{2(p-3)}{\tau}} - \frac{\ln\tau}{2w\tau}
+O(\tau^{-1})
\\
\displaystyle\hfill
  \mbox{ for $p>3$,}
  \end{array}\right.
\eqlabel{asympFH}
\end{equation}
where
\begin{align}
r_j(p) =& 
-\frac{1}{2}\ln\left[w^2(3-p)^3(1+p)^3(p-1)/576\right]
\eqlabel{rjp}
    -\frac{3}{2}
        +
      \frac{w(p+1)(3\Delta_j-2p^2+8p-10)}{4(3-p)}.
\end{align}
To obtain Eq.~\eq{asympFH}, we have used Eq.~\eq{piHmid} for $0\leq
p<1$, and we have combined Eqs.~\eq{piHmid} and \eq{piHlow} for
$1<p<3$, and Eqs.~\eq{piHlow} and \eq{piHhi} for $p>3$. Only the case
$p\geq 0$ is given by Eq.~\eq{asympFH}. For $p<0$, one can use that by
the definition Eq.~\eq{defFH}, $f_j^\sH(p;\tau)$ is an odd function of
$p$.

We remark that as we approach the limits of the range of $p$ values
for which each of the expressions for the PDF $\pi_j^\sH$,
[i.e. Eqs.~\eq{piHlow}, \eq{piHmid} and \eq{piHhi}] holds, their
r.h.s.'s diverge. Similarly, the r.h.s.'s of Eq.~\eq{asympFH} diverge
there.  These divergences are an artifact of the expansion in inverse
powers of $\tau$, as can be seen because such divergences are not
observed in Sec.~\ref{Analytic} (i.e., in Figs.~\ref{fig:4a} and
\ref{fig:5}). In fact, for any $\tau$, {\em Mathematica} shows us that
$q^*$ is a continuous and differentiable function of $p$\cite{fnfour}
(with a discontinuity in the derivative $\partial q^*/\partial p$ only
as $\tau\to\infty$), leading to a continuously varying
$\pi^\sH_j(p;\tau)$ with $p$. Furthermore, considering the regions of
$p$ values in which Eq.~\eq{asympFH} differs (due to these
divergences) noticeably from the full saddle-point approximation, one
finds from {\it Mathematica} that these regions shrink to zero as
$\tau\to\infty$.

For $\tau\to\infty$, the results in Eq.~\eq{asympFH} become
\begin{equation}
  \lim_{\tau\to\infty} f_j^\sH(p;\tau) = \left\{\begin{array}{ll}
  p&\mbox{ for $0\leq p<1$}\\
  p-(p-1)^2/4&\mbox{ for $1<p<3$}\\
  2&\mbox{ for $p>3$}
  \end{array}\right.
\eqlabel{infiniteasympFH}
\end{equation}
This is the extension of the conventional FT as it holds for infinite
time.  Comparing with Eq.~\eq{preciseHFT}, it is clear that the
conventional FT only holds for not too large fluctuations,
i.e. $|p|<1$. For larger $p$, there is first a quadratic deviation
after which the ratio between the probability for positive and for
negative fluctuations becomes a constant. In contrast, for the
conventional FT, $f_j^\sH$ keeps increasing linearly with the size of
the fluctuation $p$.

Equation \eq{asympFH} gives the extension of the fluctuation theorem
for {\em finite} $\tau$. The deviations from the infinite $\tau$
results are of a different character for small ($|p|<3$) and large
($|p|>3$) fluctuations. The $f_j^\sH(p;\tau)$ for large positive
fluctuations ($p>3$) approaches $2$ with corrections that scale as
$1/\sqrt{\tau}$. But whereas in the infinite $\tau$ limit in
Eq.~\eq{infiniteasympFH}, the function no longer grows above $2$
beyond $p=3$, in the Eq.~\eq{asympFH} for finite $\tau$ this function
keeps growing to leading order as a square root, i.e.,
$f_j^\sH(p;\tau)\sim\sqrt{(p-3)/\tau}$.  The prefactor of
$\sqrt{(p-3)/\tau}$ depends on whether we consider the transient or
the stationary case --- it is multiplied by $\sqrt{2}$ for the former
and by $2\sqrt2$ for the latter --- but it does not depend on $w$. For
small fluctuations, on the other hand, which might be more relevant
for many applications, we see corrections to the infinite $\tau$ limit
of $O(\tau^{-1})$, which do depend on $w$. The value of $w$ determines
whether the curve starts around $p=0$ above or below the line with
slope $1$, and thus also whether the slope of $1$ is approached from
below or from above.  In fact, expanding Eq.~\eq{asympFH} around
$p=0$, we get
\begin{equation}
  f_{j}^\sH(p;\tau) =
  \left[1+\frac{2}{\tau}\left(\frac{4-3\Delta_j}{9}-\frac{1}{w}\right)\right]
  p + O(p^2).
\eqlabel{slope}
\end{equation}
This shows that the critical value for $w$ is $w_c=9/(4-3\Delta_j)$,
i.e., $w_c=9$ for the transient and $w_c=9/4$ for the stationary
case. For $w=w_c$, the slope (near $p=0$) is $1$ for finite times (up
to corrections of order $\tau^{-2}$), while the slope of $1$ is
approached for large $\tau$ from above for $w>w_c$, and from below for
$w<w_c$.  In contrast, for the conventional TFT, the slope is $1$ for
all $\tau$, while for the conventional SSFT, at least for the work
done on the system, the slope always approaches $1$ from above
irrespective of $w$ with increasing $\tau$.

\section{Discussion} 
\label{Discussion}   

1. This paper treats fluctuations in the heat developed in a system of
a Brownian particle in water, confined by a harmonic potential, which
moves at constant velocity through the fluid, dragging the Brownian
particle with it.  The theory of heat fluctuations developed in this
system was based on an overdamped Langevin equation for the position
of the particle.  This theory required a far more sophisticated
analysis than was used in the previous paper\cite{VanZonCohen02b} for
work fluctuations. It should be mentioned that some of this same
sophistication is also found in the work of Farago for a quantity
different from both the work and the heat\cite{Farago02}.

2. In essence, our theory deals with the fluctuations of the
quantities occurring in the first law of thermodynamics, i.e., work,
heat and internal energy.  The energy balance for the system is
\begin{equation}
  Q_\tau = W_\tau - \Delta U_\tau.
\eqlabel{lastt}
\end{equation}
Here $W_\tau$ is the total work done on the system during a time
$\tau$, $Q_\tau$ is the heat produced by the Brownian particle in
water in that time, and $\Delta U_\tau$ is the difference in potential
energy of the particle in the harmonic potential in the same time
interval. Equation \eq{lastt}, which is basically the first law, can
be applied both to averages as well as to fluctuations, because it
expresses energy conservation, which holds both macroscopically and
microscopically.

3.  The theory gives extensions of the conventional SSFT and TFT in
Eq.~\eq{asympFH}. In the limit $\tau\to\infty$, leading to
Eq.~\eq{infiniteasympFH}, this new theorem coincides with the
conventional TFT and SSFT only for small fluctuations $p<1$ (as
Rey-Bellet and Thomas also found for a different
system\cite{ReyBelletThomas02}), while for larger fluctuations the
behavior is completely different from the conventional ones. We will
now explain the qualitative behavior of work and heat fluctuations.
For that, it is useful first to consider, in point 4 below, the system
in equilibrium, i.e., in the situation in which the harmonic confining
potential does not move. In point 5, we then discuss the
non-equilibrium case.

4. In equilibrium, work, potential energy and heat behave as
follows.~(i)~Work: There is no displacement and hence no work is
done. Therefore the PDF of $W_\tau$ is a delta-function:
$P(W_\tau)=\delta(W_\tau)$. (ii)~Potential: The PDF of the potential
energy of the particle in the harmonic potential, $U$, is given by a
Boltzmann factor $\sim\exp[-U]$ ($k_BT=1$), since the particle in the
harmonic potential can be seen as a subsystem of a larger one.  From
this, one sees that the PDF of $\Delta U_\tau$ behaves for large
$\tau$ similarly in its tails, i.e., $P(\Delta
U_\tau)\sim\exp(-|\Delta U_\tau|)$. (iii)~Heat:~For the PDF of the
heat fluctuation, using Eq.~\eq{lastt} with $W_\tau=0$, we see that
$Q_\tau$ also has exponential tails: $P(Q_\tau)\sim\exp(-|Q_\tau|)$.

5. We now turn to the non-equilibrium, stationary state
case. (i)~Work:~ the PDF of $W_\tau$ is now a
Gaussian\cite{VanZonCohen02b}.  (ii)~Potential:~The PDF of $\Delta
U_\tau$ is expected to have still exponential tails, at least near
equilibrium, i.e., $P(\Delta U_\tau)\sim\exp(-|\Delta U_\tau|)$.
(iii)~Heat:~The effect of the interplay between $W_\tau$ and $\Delta
U_\tau$ on the behavior of heat fluctuations can be deduced using
Eq.~\eq{lastt}.  We have to separately consider small and large
fluctuations.  For large $\tau$, $W_\tau$ and $Q_\tau$ grow on average
linearly in time, while $\Delta U_\tau$ stays of $O(1)$. Hence, for
small fluctuations of these quantities (near their averages) one can
neglect $\Delta U_\tau$ in Eq.~\eq{lastt}, so that $Q_\tau\approx
W_\tau$, and the behavior of $Q_\tau$ is very similar behavior to that
of $W_\tau$, i.e. Gaussian-like.  On the other hand, when a large
fluctuation of $Q_\tau$ occurs, it is less likely to be due to a large
fluctuation of $W_\tau$ than to a large fluctuation of $\Delta
U_\tau$.  This is so because the tails of the Gaussian PDF for
$W_\tau$ are much smaller than the exponential tails of the PDF for
$\Delta U_\tau$.  As a result, $W_\tau$ will be near its average while
$\Delta U_\tau$ is large. For the sake of the argument, we put
$W_\tau$ equal to its average, $\average{W_\tau}$, which coincides
with $\average{Q_\tau}$ for large $\tau$. Hence by Eq.~\eq{lastt},
$Q_\tau\approx \average{W_\tau} - \Delta U_\tau= \average{Q_\tau} -
\Delta U_\tau$.  Using that the PDF for $\Delta U_\tau$ behaves as
$\exp(-|\Delta U_\tau|)$ in its tails, we see that the PDF for
$Q_\tau$ has tails of the form $\exp(-|Q_\tau-\average{Q_\tau}|)$, in
agreement with Eqs.~\eq{piHlow} and \eq{piHhi}.

6.  We saw in Ref.~\cite{VanZonCohen02b} that the work obeys the
conventional FT in the limit $\tau\to\infty$. Heat and work
fluctuations are expected to behave similarly for small fluctuations
(cf. point 5), and therefore the conventional FT is obeyed also by the
heat fluctuations for $\tau\to\infty$ for {\em small} enough
fluctuations.  However, for {\em large} fluctuations, we get a
different behavior.  Using
$P(Q_\tau)\sim\exp(-|Q_\tau-\average{Q_\tau}|)$, the fluctuation
function $f_j^\sH$ which is defined by Eq.~\eq{defFH} and can be
written as $\ln[P(Q_\tau)/P(-Q_\tau)]/\average{Q_\tau}$ becomes
$(-|Q_\tau-\average{Q_\tau}|+
|Q_\tau+\average{Q_\tau}|)/\average{Q_\tau} =2$. This explains
qualitatively the behavior expressed in Eq.~\eq{infiniteasympFH}.

7. The symmetry relation in Eq.~\eq{TFTq} is very reminiscent of the
one used by Lebowitz and Spohn\cite{LebowitzSpohn99} in their work on
the fluctuation theorem (i.e., $e(\lambda)=e(1-\lambda)$), although
their method of large deviation theory allows only a treatment of the
behavior of $P^\sH_j$ and $f_j^\sH$ for $\tau\to\infty$. The precise
connection between their models (and the conventional FT they find),
and our model (and the extended FTs) is not clear.

8. We derived the extended infinite-$\tau$ FT of
Eq.~\eq{infiniteasympFH} already in Ref.~\cite{VanZonCohen03a} using
large deviation theory. In fact, the saddle-point method reduces to
the large deviation theory of that paper in the limit
$\tau\to\infty$. We will not prove this here, but remark that if one
takes $\tau\to\infty$ in Eq.~\eq{gS}, then one gets $-q(i-q)$. Setting
$q=i\lambda$, this becomes $\lambda(1-\lambda)$, which is the form of
the quantity $e(\lambda)$ used in the large deviation theory in
Ref.~\cite{VanZonCohen03a} for $|\lambda|<1$.  The expression breaks
down when the correction terms to $-q(i-q)$ in Eq.~\eq{gS} become
infinite, i.e. at $\lambda=\pm 1$. This restriction of $|\lambda|<1$
was also crucial in obtaining the extended FT in
Ref.~\cite{VanZonCohen03a}.  We remark that likewise, in
Fig.~\ref{fig:3}, the singularities at $q=\pm i$ restrict the saddle
point to the region $|q|<1$ as well , and that this in turn leads to
exponential tails of $P(Q_\tau)$, which finally give rise to the
extensions of the heat FT.

9. One of the important and striking results of the extended FTs is
that the probability ratio for negative to positive fluctuations in
the heat production by the Brownian particle is much larger than that
given by the conventional FTs.

\section*{\uppercase{Acknowledgments}}

This work has been supported by the Office of Basic Engineering of the
US Department of Energy, under grant No. DE-FG-02-88-ER13847.

\appendix

\section{PDF of work over a time interval and endpoint positions}
\label{appA}

Here, we will determine the Gaussian joint distributions $P^*_j$ of
the work $W_\tau$ over a time interval of length $\tau$ from time $t$
to $t+\tau$ and the positions of the Brownian particle at the
beginning, ${\Delta\mathbf{x}}_t$, and the end,
${\Delta\mathbf{x}}_{t+\tau}$, where
\begin{equation}
  {\Delta\mathbf{x}}_t ={\mathbf{x}}_t - \mathbf{v}^* t.
  \eqlabel{defdxt}
\end{equation}
These are needed for the numerical sampling method in
Sec.~\ref{Numerical} as well as in the calculating of the Fourier
transform of $P^\sH_j$ in Eq.~\eq{Fourierint} which is evaluated in
App.~\ref{appB}. We are interested both in the transient case, for
which $j=\mathrm T$ and $t=0$, and in the stationary state, for which
$j=\mathrm S$ and $t\to\infty$.

For notational convenience, we introduce a seven dimensional vector
$\mathbf{a} = (W_\tau,{\Delta\mathbf{x}}_1,{\Delta\mathbf{x}}_2)$. The
PDF $P^*_j$ is characterized by the moments
\begin{equation}
  \bar{\mathbf{a}}_j 
    =
	 \int \!\!d \mathbf{a}\, P^*_j(\mathbf{a};\tau) \mathbf{a},
\eqlabel{A1}
\end{equation}
which is a seven dimensional vector, and
\begin{align}
  \mathsf{A} &= \int \!\!d \mathbf{a}\, P^*_j(\mathbf{a};\tau)
    (\mathbf{a}-\bar{\mathbf{a}}_j)
    (\mathbf{a}-\bar{\mathbf{a}}_j)^\dagger, \eqlabel{A2}
\end{align}
Here the superscript dagger ($\dagger$) denotes the transpose.  It is
clear from this definition that $\mathsf{A}$ is a real symmetric
$7\times7$ matrix. Also, it has $\det \mathsf{A}\geq 0$.

Once these moments are known,
$P^*$ is given by:
\begin{equation}
  P^*_j(\mathbf{a};\tau) 
    = 
	\frac{
	  e^{-\frac12(\mathbf{a}-\bar{\mathbf{a}}_j)^\dagger\cdot\mathsf{A}^{-1}
	     \cdot(\mathbf{a}-\bar{\mathbf{a}}_j)}
	}{\sqrt{\det(2\pi\mathsf{A})}}.
\eqlabel{A3}
\end{equation}
Note that if $\det \mathsf{A}=0$, then the PDF $P^*$ is a delta
function (in one or more directions).

To give the specific form for $\mathbf{a}$ and $\mathsf{A}$, we first write
\begin{equation}
\bar{\mathbf a}_j = \begin{pmatrix}
  \bar{\mathrm a}^{(1)}_j\\\bar{\mathbf a}^{(2)}_j\\\bar{\mathbf a}^{(3)}_j
		\end{pmatrix},
\eqlabel{A31}
\end{equation}
(where $\bar{\mathrm a}_j^{(1)}$ is a scalar and $\bar{\mathbf
a}^{(2)}_j$ and $\bar{\mathbf a}^{(3)}_j$ are three-vectors) and
\begin{equation}
  \mathsf{A} 
    =
	\begin{pmatrix}
		A_{11}&\mathbf{A}_{21}^\dagger&\mathbf{A}_{31}^\dagger\\
	       	\mathbf{A}_{21}&\mathsf{A}_{22}&\mathsf{A}_{32}^\dagger\\
	       	\mathbf{A}_{31}&\mathsf{A}_{32}&\mathsf{A}_{33}
	\end{pmatrix},
\eqlabel{A5}
\end{equation}
(where $A_{11}$ is a scalar, $\mathbf{A}_{21}$ and $\mathbf{A}_{31}$
are three-vectors, $\mathsf{A}_{22}$, $A_{23}$ and $\mathsf{A}_{33}$
are $3\times3$ matrices).

The specific form for $\bar{\mathbf{a}}_j$ for the transient case,
i.e., $\bar{\mathbf{a}}_\sT$ is obtained using Eqs.~\eq{xpdf},
\eq{avWT}, \eq{defdx1T}, \eq{defdx2T}, \eq{defdxt}, \eq{A1} and
\eq{A31}, which yield
\begin{align}
   \bar{\mathrm a}^{(1)}_\sT &= \average{W_\tau}_\sT =w ( \tau - 1 + e^{-\tau})
\eqlabel{A4a}
\\
\bar{\mathbf a}^{(2)}_\sT &= \average{{\Delta\mathbf{x}}_0} ={\mathbf 0}
\eqlabel{A4b}
\\
\bar{\mathbf a}^{(3)}_\sT &= \average{{\Delta\mathbf{x}}_\tau} =(e^{-\tau}-1)\mathbf{v}^*.
\eqlabel{A4}
\end{align}
The sub-elements of $\mathsf{A}$ for the transient case are explicitly
determined from Eqs.~\eq{avX}, \eq{avXtX}, \eq{Wdef}, \eq{varWT},
\eq{defdx1T}, \eq{defdx2T}, \eq{defdxt}, \eq{A1} and
\eq{A31}--\eq{A4}, giving
\begin{align}
A_{11} &=
        \laverage{[W_\tau-\average{W_\tau}]^2}_\sT
=
	2  w (\tau - 1+e^{-\tau})
\eqlabel{A6}
\\
\mathbf{A}_{21} &=
        \laverage{[{\Delta\mathbf{x}}_0-\average{{\Delta\mathbf{x}}_0}][W_\tau-\average{W_\tau}]}
\nonumber\\&
	=  (e^{-\tau}-1) \mathbf{v}^*
\eqlabel{A7}
\\
\mathbf{A}_{31}&=
      \laverage{[{\Delta\mathbf{x}}_\tau-\average{{\Delta\mathbf{x}}_\tau}][W_\tau-\average{W_\tau}]}
\nonumber\\
	&= (e^{-\tau}-1) \mathbf{v}^* 
\eqlabel{A8}\\
\mathsf{A}_{22} &=  
      \laverage{[{\Delta\mathbf{x}}_0-\average{{\Delta\mathbf{x}}_0}][{\Delta\mathbf{x}}_0-\average{{\Delta\mathbf{x}}_0}]^\dagger}
=\mathbbm{1}
\eqlabel{A9}\\
\mathsf{A}_{33} &= 
      \laverage{[{\Delta\mathbf{x}}_\tau-\average{{\Delta\mathbf{x}}_\tau}][{\Delta\mathbf{x}}_\tau-\average{{\Delta\mathbf{x}}_\tau}]^\dagger}
=
\mathbbm{1}
\eqlabel{A10}\\
\mathsf{A}_{32}  &= 
      \laverage{[{\Delta\mathbf{x}}_\tau-\average{{\Delta\mathbf{x}}_\tau}][{\Delta\mathbf{x}}_0-\average{{\Delta\mathbf{x}}_0}]^\dagger}
=
e^{-\tau}\mathbbm{1}
\eqlabel{A11}
\end{align}

For the stationary case, the specific forms for the components of
$\bar{\mathbf{a}}_\sS$ are found from Eqs.~\eq{xpdf}, \eq{avWS},
\eq{defdx1SS}, \eq{defdx2SS}, \eq{defdxt}, \eq{A1} and \eq{A31}:
\begin{align}
\bar{\mathrm a}^{(1)}_\sS &=\average{W_\tau}_\sS = w \tau
\eqlabel{A12a}\\
\bar{\mathbf a}^{(2)}_\sS &= \lim_{t\to\infty}\average{{\Delta\mathbf{x}}_t} = - \mathbf{v}^*
\eqlabel{A12b}\\
\bar{\mathbf a}^{(3)}_\sS &= \lim_{t\to\infty}\average{{\Delta\mathbf{x}}_{t+\tau}} = - \mathbf{v}^*
\eqlabel{A12}
\end{align}
while the sub-elements of $\mathsf{A}$ are in that case, by
Eqs.~\eq{defdx1SS}, \eq{defdx2SS}, \eq{A1}, \eq{A31} and \eq{A5},
\begin{align}
A_{11} &=
  \lim_{t\to\infty}      \laverage{[W_\tau-\average{W_\tau}]^2}
=        \laverage{[W_\tau-\average{W_\tau}]^2}_\sS
\eqlabel{A13}
\\
\mathbf{A}_{21} &=
\lim_{t\to\infty}\laverage{[{\Delta\mathbf{x}}_t-\average{{\Delta\mathbf{x}}_t}][W_\tau-\average{W_\tau}]}
\eqlabel{A14}
\\
\mathbf{A}_{31}&=
      \lim_{t\to\infty}
\laverage{[{\Delta\mathbf{x}}_{t+\tau}-\average{{\Delta\mathbf{x}}_{t+\tau}}][W_\tau-\average{W_\tau}]}
\eqlabel{A15}\\
\mathsf{A}_{22} &=  
 \lim_{t\to\infty}
\laverage{[{\Delta\mathbf{x}}_t-\average{{\Delta\mathbf{x}}_t}][{\Delta\mathbf{x}}_{t}-\average{{\Delta\mathbf{x}}_{t}}]^\dagger}
\eqlabel{A16}\\
\mathsf{A}_{33} &=  \lim_{t\to\infty}
\laverage{[{\Delta\mathbf{x}}_{t+\tau}-\average{{\Delta\mathbf{x}}_{t+\tau}}][{\Delta\mathbf{x}}_{t+\tau}-\average{{\Delta\mathbf{x}}_{t+\tau}}]^\dagger}
\eqlabel{A17}\\
\mathsf{A}_{32}  &= \lim_{t\to\infty}
\laverage{[{\Delta\mathbf{x}}_{t+\tau-}\average{{\Delta\mathbf{x}}_{t+\tau}}][{\Delta\mathbf{x}}_t-\average{{\Delta\mathbf{x}}_t}]^\dagger}
\eqlabel{A18},
\end{align}
which turn out to be identical to those of the transient case in
Eqs.~\eq{A6}-\eq{A11} upon direct evaluation using
Eqs.~\eq{avX}--\eq{Wdef}, \eq{avWS}, \eq{defdxt} and
\eq{A12a}--\eq{A12}.  This is why we did not denote a $j$ dependence
of $\mathsf{A}$.

\section{Fourier transform of $\pi_j^\sH$}
\label{appB}

The Fourier transform of the PDF of heat will be calculated here,
starting from Eq.~\eq{Fourierint}.  To calculate the $\hat
P_j(q;\tau)$ from that equation, we define the quantities
\begin{eqnarray}
  \mathbf{c} 
     &=& \left(\begin{array}{c}1\\0\\0\end{array}\right)
\eqlabel{1.46}
\\
  \mathsf{B} 
    &=& 
	 \begin{pmatrix}
		0&0&0\\
		0&\mathbbm{1}&0\\
		0&0&-\mathbbm{1}
		\end{pmatrix},
\eqlabel{1.47}
\end{eqnarray}
so that one can write in the exponent in Eq.~\eq{Fourierint} as
$W_\tau-\sfrac{1}{2}\left\{|{\Delta\mathbf{x}}_2|^2-|{\Delta\mathbf{x}}_1|^2\right\}
=
\mathbf{c}\cdot\mathbf{a}+\sfrac12\mathbf{a}^\dagger\cdot\mathsf{B}\cdot\mathbf{a}$.
Using Eq.~\eq{A3}, one obtains then
\begin{equation}
  \hat P_j(q;\tau) 
    = 
	\int \!\! d \mathbf{a}\,
	\frac{e^{-\frac12(\mathbf{a}-\bar{\mathbf{a}}_j)^\dagger\cdot\mathsf{A}^{-1}\cdot(\mathbf{a}-\bar{\mathbf{a}}_j)
		+\frac12iq\mathbf{a}^\dagger\cdot\mathsf{B}\cdot\mathbf{a}+iq\mathbf{c}\cdot\mathbf{a}}}
	{\sqrt{\det(2\pi\mathsf{A})}}
.
\eqlabel{1.49}
\end{equation}
To evaluate this, the exponent is first rewritten as
\begin{align}
  -\frac12(\mathbf{a}-\bar{\mathbf{a}}_j)^\dagger\cdot\mathsf{A}^{-1}\cdot(\mathbf{a}-\bar{\mathbf{a}}_j)
			+\frac12iq\mathbf{a}^\dagger\cdot\mathsf{B}\cdot\mathbf{a}+iq\mathbf{c}\cdot\mathbf{a}
\nonumber\\
  = -\frac12(\mathbf{a}-\mathbf{a}'_j)^\dagger\cdot\left[ \mathsf{A}^{-1}-iq\mathsf{B}\right]\cdot(\mathbf{a}-\mathbf{a}'_j) 
+ d_j,
\eqlabel{1.50}
\end{align}
where $ \mathbf{a}'_j = \left[\mathsf{I}-iq\,\mathsf{A}\cdot\mathsf{B}
\right]^{-1}\cdot(\bar{\mathbf{a}}_j+iq\,\mathsf{A}\cdot\mathbf{c})$
and
\begin{equation}
  d_j
    = 
	\frac{iq}{2}\Big[
	(\mathsf{B}\cdot\bar{\mathbf{a}}_j+\mathbf{c})^\dagger\cdot(\mathsf{I}-iq\,\mathsf{A}\cdot\mathsf{B})^{-1}\cdot
	(\bar{\mathbf{a}}_j+iq\,\mathsf{A}\cdot\mathbf{c})
	+\bar{\mathbf{a}}_j\cdot\mathbf{c}
	\Big].
\eqlabel{1.56} 
\end{equation}
Here, $\mathsf{I}$ is the $7\times7$ identity matrix.  Then
substituting Eq.~\eq{1.50} in Eq.~\eq{1.49} and changing the
integration variable to ${\mathbf{x}}=\mathbf{a}-\mathbf{a}'_j$ yields
\begin{align} 
  \hat P_j(q;\tau)
    &= 
	\int \!\!d {\mathbf{x}}\,
	e^{-\frac12{\mathbf{x}}^\dagger\cdot(\mathsf{A}^{-1}-iq\mathsf{B})\cdot{\mathbf{x}}}
	\frac{e^{d_j}}{\sqrt{\det(2\pi\mathsf{A})}}
\eqlabel{theint}
\\
    &=
	\frac{e^{d_j}}{\sqrt{\det(\mathsf{I}-iq\,\mathsf{A}\cdot\mathsf{B})}}
\eqlabel{A26},
\end{align}
where the identity
$\det(\mathsf{A})\det(\mathsf{B})=\det(\mathsf{A}\mathsf{B})$ has been
used.  To make Eq.~\eq{A26} into an explicit expression for $\hat
P_j$, the inverse of the matrix $(\mathsf{I} - iq
\mathsf{A}\cdot\mathsf{B})$ is required in the expression for $d_j$ in
Eq.~\eq{1.56}, and in Eq.~\eq{A26}, its determinant.  These are
obtained as follows. Using Eqs.~\eq{A5}, \eq{A6}-\eq{A11} and
\eq{1.47}, it follows that
\begin{equation}
\mathsf{I}-iq\,\mathsf{A}\cdot\mathsf{B} =
\begin{pmatrix}
	1 &iq (e^{-\tau}-1)\mathbf{v}^{*\dagger}  & iq(e^{-\tau}-1)\mathbf{v}^{*\dagger}\\
	{\mathbf 0} &(1-iq)\mathbbm{1}             & iq e^{-\tau}\mathbbm{1}\\
	{\mathbf 0} &-iq e^{-\tau}\mathbbm{1}      & (1+iq)\mathbbm{1}
	\end{pmatrix}.
\eqlabel{1.57} 
\end{equation}
The determinant of this matrix is
\begin{equation}
\det(\mathsf{I}-iq\,\mathsf{A}\cdot\mathsf{B}) = \left[1+(1-e^{-2\tau})q^2\right]^3.
\eqlabel{1.58}
\end{equation}
For the inverse of Eq.~\eq{1.57}, we get
\begin{align}
&\left(	\mathsf{I} -iq\,\mathsf{A}\cdot\mathsf{B}\right)^{-1} =
\nonumber\\
&
\begin{pmatrix}
1 &
\frac{iq(e^{-\tau}-1)[1+iq(1-e^{-\tau})]\mathbf{v}^{*\dagger}}{1+(1-e^{-2\tau})q^2} &
\frac{iq(1-e^{-\tau})[1-iq(1-e^{-\tau})]\mathbf{v}^{*\dagger}}{1+(1-e^{-2\tau})q^2} \\
{\mathbf 0} &
\frac{1+iq}{1+(1-e^{-2\tau})q^2}\mathbbm{1} &
\frac{-iqe^{-\tau}}{1+(1-e^{-2\tau})q^2}\mathbbm{1} \\
{\mathbf 0} &
\frac{iq e^{-\tau}}{1+(1-e^{-2\tau})q^2}\mathbbm{1} &
\frac{1-iq}{1+(1-e^{-2\tau})q^2}\mathbbm{1}
\end{pmatrix}.
\eqlabel{1.64}
\end{align}
We now have the material needed to calculate $d_j$ from Eq.~\eq{1.56}
explicitly.  To calculate Eq.~\eq{1.56}, we use Eqs.~\eq{A31}, \eq{A5},
\eq{A6},\eq{A7},\eq{A8}, \eq{1.46},\eq{1.47} and \eq{1.64}, to find,
after some rearrangements,
\begin{widetext}
\begin{align}
d_j= 
iq\bigg\{
&
	\frac{1}{1+(1-e^{-2\tau})q^2}\Big[
	  -iq^3w(1-e^{-\tau})^3
\Big.
	-iq e^{-\tau}\bar{\mathbf a}^{(2)}_j\cdot\bar{\mathbf a}^{(3)}_j
	+ \sfrac12 |\bar{\mathbf a}^{(2)}_j|^2(1+iq)
	- \sfrac12|\bar{\mathbf a}^{(3)}_j|^2(1-iq)
\nonumber\\&
\Big.
	-   iq  (1-e^{-\tau})[1+iq(1-e^{-\tau})] \mathbf{v}^*\cdot\bar{\mathbf a}^{(2)}_j
\Big.
	+   iq  (1-e^{-\tau})[1-iq(1-e^{-\tau})]  \mathbf{v}^*\cdot\bar{\mathbf a}^{(3)}_j
	\Big]
\nonumber\\&
+	 \bar{\mathrm{a}}^{(1)}_j+iq w (\tau-1+e^{-\tau})
	\bigg\}
\eqlabel{2.15}
\end{align}
\end{widetext}
Furthermore, from Eqs.~\eq{A4a}--\eq{A4} and \eq{2.15}, $d_\sT$ for the
transient case follows as
\begin{equation}
d_\sT =
wq(i-q)\left\{\tau-\frac{[1-e^{-\tau}][1+(\sfrac12+2q^2)(1-e^{-\tau})]}
{1+(1-e^{-2\tau})q^2}\right\}
\eqlabel{1.71}
\end{equation}
while using Eqs.~\eq{A12a}--\eq{A12} and \eq{2.15}, $d_{\sS}$ for the
stationary case, is, after some rewriting, found to be
\begin{equation}
d_\sS =
  w q(i-q)\left\{\tau
	 -  \frac{2q^2(1-e^{-\tau})^2}{1+q^2(1-e^{-2\tau})}
\right\}.
\eqlabel{2.29}
\end{equation}

\noindent
Finally, by Eqs.~\eq{A26} and \eq{1.58}, these expressions for $d_j$
yield for the Fourier transforms explicitly
\begin{align}
  \hat P_{\sT}(q;\tau) &=
  \frac{
e^{
wq(i-q)\left\{\tau-\frac{[1-e^{-\tau}][1+(1/2+2q^2)(1-e^{-\tau})]}
{1+(1-e^{-2\tau})q^2}\right\}
}
}
       {[1+(1-e^{-2\tau})q^2]^{3/2}}
\\
  \hat P_{\sS}(q;\tau) &=
  \frac{
e^{
wq(i-q)\left[\tau-\frac{2q^2(1-e^{-\tau})^2}{1+(1-e^{-2\tau})q^2}\right]
}
}
       {[1+(1-e^{-2\tau})q^2]^{3/2}}
.
\end{align}


\begin{thebibliography}{30}
\bibitem{fnone}
The dynamical systems are required to have an isoenergetic Gaussian
thermostat for this to be strictly true, otherwise correction terms
appear, though these might possibly vanish in the large time or system
size limit.
\bibitem{fntwo}
Also, the integrated TFT and SSFT
exist\cite{Wangetal02,VanZonCohen02b}, which we do not discuss here
since they are constructed only to improve the statistics in
experiments or simulations.
\bibitem{fnthree}
In Eq.~\eq{Qdef} the possibility of energy being stored as kinetic
energy of the Brownian particle has been neglected. While this in
principle should be included, fluctuations in velocity decay on a time
scale $\ll 1$ in the overdamped case, so that we may neglect them on
time scales of $O(1)$.
\bibitem{fnfive}
We have verified using the numerical Fast Fourier Transform inverse of
Sec.~\ref{Numerical}, that neglecting the exponential terms in the
Fourier transform $\hat P_j$ in Eq.~\eq{hpt} has negligible
consequences for $\pi_j^\sH$ in Eq.~\eq{inverse} if $\tau\gtrsim 3$.
\bibitem{fnsix}
With exponential terms neglected in Eq.~\eq{hpt}, $P_j(q;\tau)$
behaves as $\exp[-(\tau-2)q^2]$ for large $q$, so that on the r.h.s.\
of Eq.~\eq{inverse} the integral for the Fourier inverse diverges for
$\tau<2$. This is an artifact of neglecting the exponentials.
\bibitem{fnfour}
It is useful to have an analytic result for the saddle points in {\em
Mathematica}'s memory because this shows that $q^*$ is a smooth
function of $p$ and allows for the derivative $\partial q^*/\partial
p$ needed in Eq.~\eq{spapprox} to be obtained as well.
\bibitem{fnnine}
Corrections to the position of the saddle point are here of higher
order. To see this, let us split up $h$ in Eq.~\eq{saddlepointmethod}
as $h_0+\delta h$, and the position of the maximum of $h$ as
$q^*=x+\delta x$.  Assuming $\delta x$ and $\delta h$ to scale as
$\tau^{-\alpha}$ and $\tau^{-\beta}$ respectively, with $\alpha,\beta
>0$, we expand $h(q^*)= h(x) + \delta h'(x) \delta x + h''_0(x) \delta
x^2/2 +\delta h''(x) \delta x^2/2 + \ldots$, using $h_0'(x)=0$.  The
terms in this expansion are $O(1)$, $O(\tau^{-\alpha-\beta})$,
$O(\tau^{-2\alpha})$ and $O(\tau^{-2\alpha-\beta})$, respectively, so
that $h(q^*)=h(x)+ O(\tau^{-\alpha-\min(\alpha,\beta)})$.
\bibitem{fneleven} 
In general, in $d$ dimensions, the corresponding expressions are found
by changing $3/2$ to $d/2$ in the denominator on the r.h.s.\ 
of Eq.~\eq{hpt} and carrying through the calculation.  For the
fluctuation function, this amounts to replacing the prefactor ``3'' of
the logarithm by $d$ in the first line of Eq.~\eq{asympFH} and to
adding prefactors $(d-1)/2$ to the logarithmic terms $\ln\tau/(w\tau)$
and $\ln\tau/(2w\tau)$ in its second and third lines, respectively. In
Eq.~\eq{rjp}, the last term remains the same, but the first two become
$-\frac{1}{2}\ln\left[w^2d^{1-d}(3-p)^d(1+p)^d(p-1)^{d-2}/64\right]$
and $-d/2$, resp. Finally, in Eq.~\eq{slope} $1/w$ becomes
$d/(3w)$.
\end{thebibliography}
\end{document}